\documentclass[12pt]{article} 

\pdfoutput=1

\usepackage{amsmath, amsfonts, amssymb}
\usepackage[margin=0.75in]{geometry}
\usepackage{graphicx}
\usepackage{hyperref}
\usepackage{setspace}
\usepackage{algorithm}
\usepackage{algpseudocode}
\usepackage{tabularx}
\usepackage{lscape} 
\usepackage{rotating}
\usepackage{hyphenat} 
\usepackage{enumitem}
\usepackage{bm}
\usepackage{placeins} 
\usepackage{setspace}
\usepackage[capitalise]{cleveref}
\usepackage{fancyhdr}
\usepackage{color}
\usepackage{adjustbox} 
\usepackage{tikz} 

\usepackage{longtable} 
\usepackage{caption}

\usepackage{array, booktabs, makecell} 

\newcommand{\edit}[1]{\color{black}#1 \color{black}}

\title{Data-driven cardiovascular flow modeling: examples and opportunities }
\author{Amirhossein Arzani$^1$ \and Scott T. M. Dawson$^2$ }

\date{}

\begin{document}

\maketitle


\begin{center}
$^1$Department of Mechanical Engineering, Northern Arizona University, Flagstaff, AZ, United States \\
$^2$Department of Mechanical, Materials and Aerospace Engineering, Illinois Institute of Technology, Chicago, IL, United States \\
\end{center}

\bigskip

\noindent Correspondence:\\
Amirhossein Arzani,\\
Northern Arizona University,\\
Flagstaff, AZ, 86011\\
Email: amir.arzani@nau.edu

\thispagestyle{empty}


\begin{abstract}

High-fidelity blood flow modeling is crucial for enhancing our understanding of cardiovascular disease. Despite significant advances in computational and experimental characterization of blood flow, the knowledge that we can acquire from such investigations remains limited by the presence of uncertainty in parameters, low resolution, and measurement noise. Additionally, extracting useful information from these datasets is challenging. Data-driven modeling techniques have the potential to overcome these challenges and transform cardiovascular flow modeling. Here, we review several data-driven modeling techniques, highlight the common ideas and principles that emerge across numerous such techniques, and provide illustrative examples of how they could be used in the context of cardiovascular fluid mechanics. In particular, we discuss principal component analysis (PCA), robust PCA, compressed sensing, the Kalman filter for data assimilation, low-rank data recovery, and several additional methods for reduced-order modeling of cardiovascular flows, including the dynamic mode decomposition (DMD), and the sparse identification of nonlinear dynamics (SINDy). All techniques are presented in the context of cardiovascular flows with simple examples. These data-driven modeling techniques have the potential to transform computational and experimental cardiovascular research, and we discuss challenges and opportunities in applying these techniques in the field, looking ultimately towards data-driven patient-specific blood flow modeling.



\end{abstract}

\newpage


\section{Introduction} \label{sec:intro}
We live in the age of data science, where the wealth of data and advances in data processing and computational power are beginning to affect every field. The field of cardiovascular fluid mechanics is no exception. Blood flow modeling has come a long way from Womersley's analytical Navier-Stokes solution for incompressible pulsatile flow in tubes during the 1950s~\cite{Womersley55} to the recent food and drug administration approval of patient-specific computational fluid dynamics (CFD) estimation of pressure drop in coronary artery disease~\cite{TaylorFonteMin13}. Advances in computational and experimental characterization of blood flow along with promising advances in the field of data science provide a unique opportunity for exciting developments in the field of cardiovascular fluid mechanics with the potential to transform our understanding of cardiovascular disease.

In modeling cardiovascular disease, blood flow and hemodynamics data are provided by multiple modalities. Patient-specific CFD, in-vivo 4D flow magnetic resonance imaging (MRI), and in-vitro particle image velocimetry (PIV) or particle tracking velocimetry (PTV) are leading modalities in providing 3D time-resolved blood flow data. However, none of these modalities are perfect. Numerical error, parameter uncertainty, imaging artifacts, low spatiotemporal resolution, and measurement noise are among the limitations of these methods.  Some open questions are summarized: How can we improve the quality and accuracy of the data generated by either of these modalities?  If we possess data from more than one modality, is it possible to generate new data with superior accuracy with respect to either dataset? How can we best extract useful information from these datasets? Given the complex spatiotemporal patterns in hemodynamics, is it possible to derive simplified representations of the data that facilitate physical understanding and further modeling?  Data-driven modeling techniques provide potential means to answer these questions. 


This work will focus on data-driven modeling techniques that seek to exploit underlying low-dimensionality and sparsity in data, and in the systems that generate such data. These methods combine classical data-driven analysis techniques originating from principal component analysis developed over a century ago by Pearson \cite{pearson1901}, with more recent advances in data-driven dynamical systems analysis \cite{Lumley1967,Holmesetal12,Schmid10,BruntonProctorKutz16}, compressive sensing \cite{candes2006compressive,donoho2006compressed,candes2006robust}, and optimal data reconstruction and interpolation \cite{barrault2004empirical,chaturantabut2010deim,manohar2018sensor}. In the class of methods that we focus on, we are typically interested in representing complex high-dimensional data in a simpler low-dimensional space, while also potentially dealing with corrupted or under-resolved data. Note that the class of data-driven modeling methods considered here share some similarities and distinctions with classical machine learning and deep learning applications. In both techniques, we are interested in learning patterns in data. In deep learning with neural networks, we are seeking optimized function approximators that are capable of revealing hidden patterns in data. Similarly, in the data-driven modeling methods discussed here, we are often interested in finding optimized models that can represent the data. A major distinction is that in what we present as data-driven modeling, we often deal with small and sometimes corrupted datasets, whereas classical deep learning techniques typically require very large datasets. Additionally, lack of interpretability is a shortcoming of current deep learning techniques, which is an important issue in fluid mechanics where we are often interested in the flow physics~\cite{Bruntonetal20}. On the other hand, data-driven reduced-order modeling facilitates physical interpretation~\cite{Tairaetal17}.


The fact that high-dimensional datasets are (at least approximately) low-rank can be observed empirically across a broad range of applications, and this ubiquitous phenomenon can also be demonstrated theoretically under certain assumptions \cite{udell2019big}. By definition, a low-rank dataset can be represented by a small number of functions. For example, low-rank data representing a snapshot in time of the state of a system can be expressed as a certain linear combination of such functions, for any time. These functions can be viewed as a small number of elements of a basis that spans the space of all possible data, whether physically realistic or not.  The fact that any physically-realistic snapshot can be represented by a small number of basis functions means that a dataset consisting of many such snapshots is sparse in this basis. The existence of a basis in which the data is sparse allows for the application of several related techniques that exploit this sparsity, such as reconstruction of full state information from a limited number of sensor measurements, which may be placed randomly, or more optimally if given more information about the system. Also, by considering data generated from a given physical system as having some type of sparsity in an appropriate basis, we can further consider notions of sparsity in the underlying equations that can generate or approximate the data. For example, if approximating a given physical system by a set of nonlinear ordinary differential equations involving a small number of state variables, the structure of the equations might typically only require a small number of nonzero terms. These nonzero terms are sparse in the space of all possible terms and can be identified from data using sparsity-promoting identification methods \cite{BruntonProctorKutz16}.

In this manuscript, we review several data-driven modeling techniques and provide examples of how they could be used in cardiovascular fluid mechanics research. The overarching goal of these techniques is to facilitate data interpretation, augment data when the data is incomplete or low-resolution, improve the quality of data, and leverage data from multiple modalities to improve data fidelity. Many of these techniques are based on a well-established, yet diverse set of mathematical tools. Familiarity with these fundamental mathematical techniques and concepts may enable cardiovascular researchers to come up with customized techniques to improve their models. The objective of this manuscript is to provide an introduction to some of these techniques with tangible examples for the cardiovascular fluid mechanics community. To help researchers who would like a more in-depth introduction to these topics, in Table~\ref{table:books},  we provide a list of excellent books that we recommend for each topic. The list is classified based on several important fundamental and applied mathematical topics that are generally important in data science and data-driven modeling for engineering and applied sciences.

\captionsetup{width=17cm}
\setlength\extrarowheight{7pt}
\begin{longtable}{ | p{0.35\textwidth} | p{0.13\textwidth} | p{0.455\textwidth} |p{0.03\textwidth} |} 
\caption{List of recommended books for an in-depth introduction to fundamental and applied mathematical topics that form the basis of data-driven modeling.} \label{table:books} \\

\hline
Book title & Author & Comments & Ref\\ \cline{1-4}
        \multicolumn{4}{| l |}{\textbf{\large{Linear algebra}}}   \\ \hline 
      Linear algebra and learning from data & G. Strang & An excellent introduction with emphasis on data-driven modeling. Author's linear algebra video lectures are available on MIT OCW website. &\hspace{-.25cm}~\cite{strang2019linear} \\ \hline

      Introduction to applied linear algebra: vectors, matrices, and least squares & S. Boyd  $\&$  L. Vandenberghe & Basic linear algebra concepts. S. Boyd's ``Introduction to Linear Dynamical Systems'' video lectures covering similar topics are available on Youtube. &\hspace{-.25cm}~\cite{boyd2018introduction} \\ \hline

     Numerical linear algebra & L. Trefethen $\&$  D. Bau & A standard introductory textbook for numerical implementation of linear algebra algorithms. & \hspace{-.25cm}~\cite{trefethen1997numerical} \\ \hline
     
        \multicolumn{4}{| l |}{\textbf{\large{Statistics}}}   \\ \hline 

    Practical Statistics for Data Scientists: 50+ Essential Concepts Using R and Python & P. Bruce et al.  & An easy read for essential statistical tools in data-driven modeling. & \hspace{-.25cm}~\cite{bruce2020practical} \\ \hline
    
    Probability and Mathematical Statistics: Theory, Applications, and Practice in R & M. Meyer & A comprehensive mathematical introduction to statistical theories. & \hspace{-.25cm}~\cite{meyer2019probability} \\ \hline

             \multicolumn{4}{| l |}{\textbf{\large{Optimization}}}   \\ \hline 

    Convex optimization & S. Boyd  $\&$  L. Vandenberghe & The standard textbook for convex optimization. S. Boyd's video lectures based on this topic are available on Youtube.  & \hspace{-.25cm} ~\cite{boyd2004convex} \\ \hline
    
                 \multicolumn{4}{| l |}{\textbf{\large{Dynamical systems}}}   \\ \hline 

   Differential equations and dynamical systems & L. Perko  & One of the best mathematical introductions to dynamical systems.  & \hspace{-.25cm}~\cite{Perko13} \\ \hline
   
Nonlinear dynamics and chaos: With applications to physics, biology, chemistry, and engineering & S. Strogatz  & A simple conceptual and practical introduction to dynamical systems. Strogatz's  video lectures based on this topic are available on Youtube.   &\hspace{-.25cm} ~\cite{strogatz2018nonlinear} \\ \hline
     
              \multicolumn{4}{| l |}{\textbf{\large{Data assimilation}}}   \\ \hline 
              
                Data assimilation: methods, algorithms, and applications & M. Asch et al.  & A rigorous and comprehensive introduction to the topic.  & \hspace{-.25cm} ~\cite{asch2016data} \\ \hline
                
                                Data assimilation: A mathematical introduction & K. Law et al.  & A mathematical introduction with Matlab codes.  & \hspace{-.25cm}~\cite{law2015data} \\ \hline

         \multicolumn{4}{| l |}{\textbf{\large{Compressed sensing and sparsity}}}   \\ \hline 
     
          Compressed sensing for engineers & A. Majumdar & An excellent practical introduction to compressed sensing with different examples such as medical imaging. & \hspace{-.25cm}~\cite{majumdar2018compressed} \\ \hline
     
              Sparse modeling: theory, algorithms, and applications & I. Rish $\&$ G.  Grabarnik   & A mathematically more detailed introduction to compressed sensing and sparsity.  & \hspace{-.25cm}~\cite{rish2014sparse} \\ \hline

 	 \multicolumn{4}{| l |}{\textbf{\large{Machine learning}}}   \\ \hline 
     
      The hundred-page machine learning book & A. Burkov & A concise but informative and conceptual introduction to machine learning. &\hspace{-.25cm} ~\cite{burkov2019hundred} \\ \hline
      
      An introduction to statistical learning & G. James et al.  & One of the most popular and accessible references in machine learning. & \hspace{-.25cm}~\cite{james2013introduction} \\ \hline
        
Matrix methods in data mining and pattern recognition & L. Eld{\'e}n  & A mathematical introduction to machine learning with emphasis on linear algebra. & \hspace{-.25cm}~\cite{elden2007matrix} \\ \hline

 	 \multicolumn{4}{| l |}{\textbf{\large{Reduced-order and data-driven modeling }}}   \\ \hline 
     
     Reduced basis methods for partial differential equations: an introduction & A.~Quarteroni~et~al. & A detailed mathematical introduction to reduced-order modeling of partial differential equations and proper orthogonal decomposition (POD). & \hspace{-.25cm}~\cite{quarteroni2015reduced} \\ \hline
     
Turbulence, coherent structures, dynamical systems and symmetry & P. Holmes et~al. & An introduction to POD with focus on applications in fluid mechanics and turbulence.  & \hspace{-.25cm} ~\cite{Holmesetal12} \\ \hline

   Dynamic mode decomposition: data-driven modeling of complex systems & N.~Kutz~et~al. & An introduction to reduced-order modeling of dynamical systems using dynamic mode decomposition (DMD).  & \hspace{-.25cm}~\cite{Kutzetal16a} \\ \hline
   
     
 Data-driven modeling \& scientific computation: methods for complex systems \& big data & N. Kutz & A comprehensive introductory book on a wide range of data-driven modeling tools and necessary mathematical preliminaries. Kutz's  video lectures based on this topic are available on Youtube.   &  \hspace{-.25cm}~\cite{kutz2013data} \\ \hline

Data-driven science and engineering: Machine learning, dynamical systems, and control & S. Brunton $\&$ N. Kutz & An excellent overview of various data-driven modeling techniques. Brunton's  video lectures based on this topic are available on Youtube.   &  \hspace{-.25cm}~\cite{brunton2019data} \\ \hline


\end{longtable}

\subsection{Manuscript organization} In each section, we present an important data-driven modeling technique. Each section is divided into subsections where we present the motivation behind  the need for the technique, the mathematical theory, example(s) relevant to cardiovascular flow problems, and a short discussion on opportunities for cardiovascular researchers as well as challenges in applying the technique to practical problems. Finally, we briefly discuss a few recent trends in computational mechanics related to data science that could be valuable in cardiovascular biomechanics modeling.

\section{Principal component analysis (PCA): detecting redundancy and low-dimensionality in data}
\subsection{Motivation} Thanks to high-performance computing, numerical simulations can provide spatiotemporally highly resolved hemodynamics data. However, often hemodynamics data possess hidden low-dimensionality, which once exposed, can reduce the dimensionality of the data and provide opportunities for advanced data analysis as discussed in the next sections. This is due to the high correlation among temporal snapshots in data (or equivalently, between different locations in space). That is, under an appropriate coordinate system,  we can reduce the size of hemodynamics data with minimal loss in accuracy. Principal component analysis (PCA) is based on singular value decomposition (SVD), which is one of the most ubiquitous and powerful matrix transformations in linear algebra~\cite{Woldetal87}. As an aside, note however that PCA (and the independently-developed Hotelling analysis \cite{hotelling1933analysis}) predated much of the theory and numerical analysis of SVD. PCA provides an optimal basis where with the minimum number of independent/uncorrelated modes one may represent the data. In other words, we seek to find a basis where the data has minimal correlation, and therefore reduced redundancy. 
What we introduce as PCA in section \ref{sec:PCA} is essentially equivalent to what is known as the proper orthogonal decomposition (POD) in fluid mechanics \cite{lumley2007stochastic}, which will be discussed explicitly in section~\ref{sec:modalTheory}. 

\subsection{Model theory and background} \label{sec:PCA} In PCA, the data is stacked into a rectangular matrix. Here, we arrange the spatial data in $n$ rows   and the temporal data in $m$ columns. Therefore, spatiotemporal hemodynamics data for a variable (e.g., velocity) could be arranged into a matrix $\mathbf{X}$

 \begin{equation}
\mathbf{X}_{(n\times m)} = \begin{bmatrix}\mathbf{W}_{1} & \mathbf{W}_{2}  & \cdots & \mathbf{W}_{m} \end{bmatrix} \;,
\label{eqn:x}
\end{equation}
where the spatial data in each snapshot (time-step) is arranged as a column. This form of data representation will frequently be used throughout the paper. Here, for clarity, we on occasion use subscripts such as $(n\times m)$ to indicate the dimensions of a matrix.  The temporal mean of the data at each spatial location  is subtracted from each row to define a new matrix $\mathbf{\tilde{X}}$


 \begin{equation}
\mathbf{\tilde{X}}_{(n\times m)} = \sqrt{\frac{1}{m}} \Big( \mathbf{X}_{(n\times m)} - \mathbf{\bar{x}}_{(n\times 1)} \mathbf{J}_{(1\times m)} \Big) \;,
\end{equation}
where  $ \mathbf{J}_{(1\times m)} =  \begin{bmatrix} 1 & 1 & \cdots & 1 \end{bmatrix}$, and $\mathbf{\bar{x}}_{(n\times 1)} =  \frac{1}{m}\sum\limits_{j=1}^{m}\mathbf{X}_{ij}$ is a column vector containing the mean of each row. Subsequently, SVD is performed on the processed data matrix

 \begin{equation}
\mathbf{\tilde{X}} = \mathbf{U}\mathbf{\Sigma} \mathbf{V}^{*} \;,
\label{eqn:SVD}
\end{equation}
where the columns of $\mathbf{U}$ and $\mathbf{V}$ are the left and right singular vectors (which are orthonormal), the diagonal values of $\mathbf{\Sigma}$ are the singular values, and $^*$ specifies the complex conjugate transpose. For the ``full'' SVD, $\mathbf{U}$ and $\mathbf{V}$ are square matrices, but here we will only need to consider the columns corresponding to nonzero singular values.

The left and right singular vectors are also eigenvectors of the equivalent space- and time-correlation matrices, satisfying
\begin{align}
\mathbf{\tilde X} \mathbf{\tilde X}^{*} \mathbf{u}_{j} &= \sigma_{j}^{2} \mathbf{u}_{j} \\
\mathbf{\tilde X}^{*}\mathbf{\tilde X} \mathbf{v}_{j} &= \sigma_{j}^{2} \mathbf{v}_{j}, \label{eq:PODsnapshots}
\end{align}
where $\mathbf{u}_{j}$ and $\mathbf{v}_{j}$ are the $j$-th columns of $\mathbf{U}$ and $\mathbf{V}$, and  $\sigma_{j}$ is the $j$-th diagonal entry of $\mathbf{\Sigma}$.

The SVD is unique (up to rotation of the singular vectors in the complex plane), and allows for an ``optimal'' low-rank approximation of the original data. More precisely, the rank-$r$ approximation $\mathbf{\tilde{X}}_{r}$ that minimizes the reconstruction error (using the Frobenius norm)
\begin{equation}
\label{eq:SVDerror}
\epsilon_r =\| \mathbf{\tilde{X}} - \mathbf{\tilde{X}}_{r}\|_F
\end{equation}
is given by 
\begin{equation}
\mathbf{\tilde{X}} _{r} = \mathbf{U}_{r}\mathbf{\Sigma}_{r} \mathbf{V}_{r}^{*},
\end{equation}
where $\mathbf{U}_{r}$ and  $\mathbf{V}_{r}$ contain the first $r$ columns of  $\mathbf{U}$ and  $\mathbf{V}$, and $\mathbf{\Sigma}_{r}$ contains the first $r$ rows and columns of $\mathbf{\Sigma}$.

Physically, with the data arranged as described in Eq.~\ref{eqn:x}, the columns of $\mathbf{U}$ denote an ordered set of spatial functions, or ``modes'', which represent the principal components of the data. The extent to which a given such function characterizes the data is given by the associated singular value (i.e.~the corresponding diagonal entry of $\mathbf{\Sigma}$).  The columns of $\mathbf{V}$ (or equivalently, rows of $\mathbf{V}^*$) give information about the relative amplitude of the corresponding spatial mode for each snapshot of data.  Time-resolved data is not required to compute the spatial PCA modes, but in the case where time-resolved data is available, these columns of $\mathbf{V}$ are functions of time. Therefore, we can equivalently express equation~\ref{eqn:SVD} as 
\begin{equation}
\mathbf{\tilde{X}} =  \sum_{j=1}^{\min(n,m)} \mathbf{u}_{j}\sigma_{j}\mathbf{v}_{j}^{*},
\end{equation}
where $\mathbf{u}_{j}$ and $\mathbf{v}_{j}$  are the $j$-th columns of $\mathbf{U}$ and $\mathbf{V}$ respectively, and $\sigma_{j}$ is the $j$-th diagonal entry of $\mathbf{\Sigma}$.

\paragraph{Remark:} For vectorial data such as velocity, the data at one time-step is still placed in one column by considering all components of the vector.


\subsection{Example: Low-dimensional behavior in cerebral and aortic aneurysm hemodynamics }
\paragraph{\textit{Problem statement:}} Do blood flow data in aneurysms exhibit low-dimensional behavior? How many modes are required to reconstruct the data? How does the complexity in cerebral and abdominal aortic aneurysm data compare?

\begin{figure}[h!]
\centering
\includegraphics[scale=0.52]{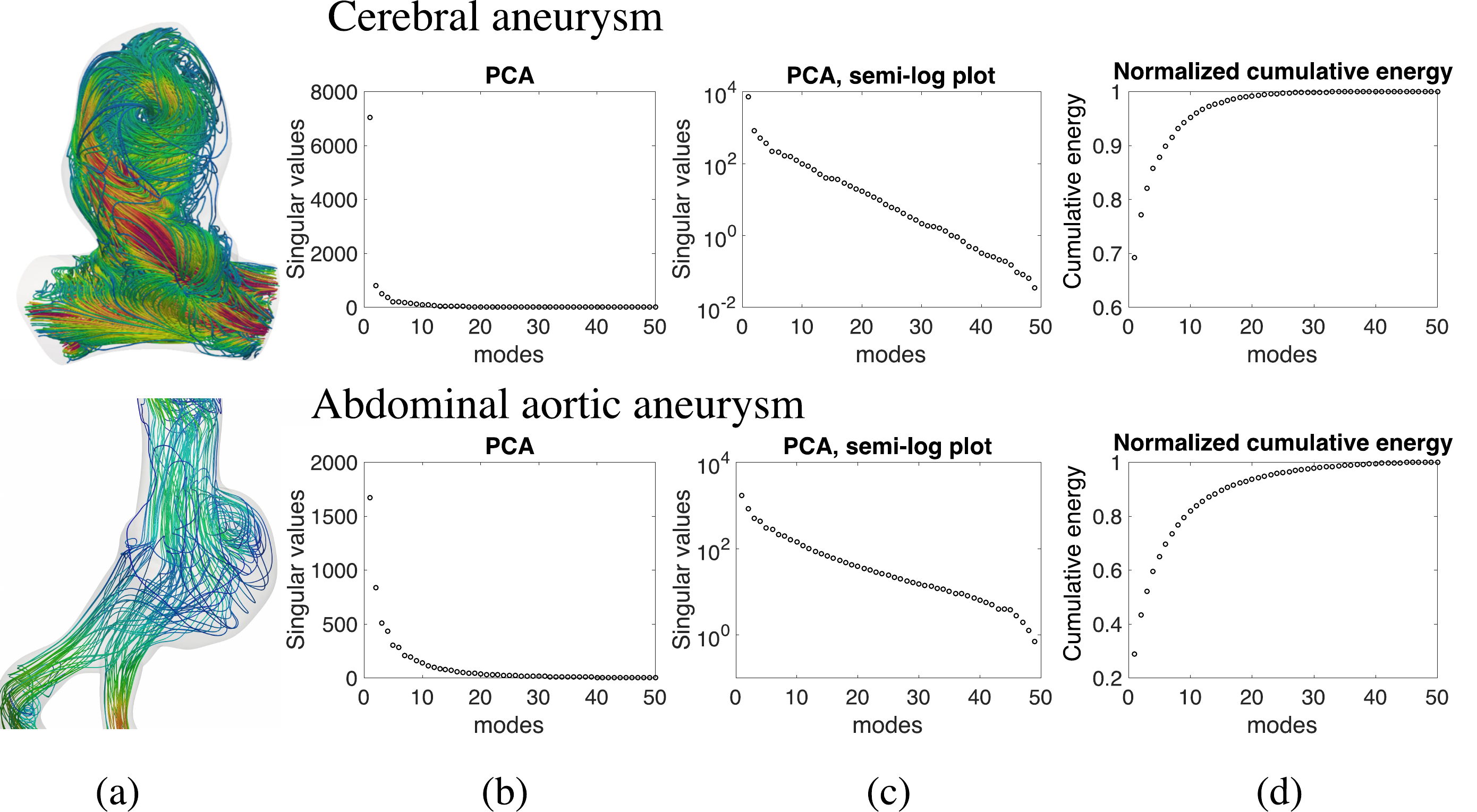}
\caption{Principal component analysis (PCA) performed on cerebral and abdominal aortic aneurysm velocity data. a) A single snapshot of the velocity streamlines is shown. b)  The singular values for different principal components (modes) are plotted. c) A semi-log plot is shown for better visualization in the range of singular values. d) The right panel shows the normalized cumulative energy defined as the cumulative sum of singular values normalized by the sum of all singular values.}
\label{fig:pca}
\end{figure}

\paragraph{\textit{Problem solution:}} Patient-specific computational fluid dynamics (CFD) results from prior work in a cerebral aneurysm~\cite{HabibiDawsonArzani20} and an abdominal aortic aneurysm (AAA)~\cite{Arzanietal14b} are used. Details about the CFD simulations can be found in~\cite{HabibiDawsonArzani20,Arzanietal14b}. 50 time-steps are used from one cardiac cycle and PCA is performed on the velocity data. The results are shown in Fig.~\ref{fig:pca}. In the cerebral aneurysm model, a single mode is capable of capturing a significant portion of the data (70\% energy) whereas more modes are required in the AAA data. To capture 90\% of the data (based on singular values), 6 and 16 modes are required for the cerebral aneurysm and AAA models, respectively. The dominant modes are capable of explaining the majority of the data. The higher number of modes in the AAA data is due to the more complex and chaotic nature of blood flow in aortic aneurysms~\cite{ArzaniShadden12}. In practice, the number of modes is selected such that a balance between model complexity and fidelity is achieved. 

\subsection{Opportunities and challenges} PCA could not only be used to post-process data but it is also a powerful pre-processing step  in various data-driven modeling tools, which will be discussed in the following sections. Herein, we have demonstrated how PCA reveals hidden low-dimensionality in hemodynamics data. In our example, the temporal data are stacked in columns, therefore low-dimensionality implies a high correlation between these columns of temporal data. Equivalently, these results also show that there must exist high amounts of correlation between measurements at different spatial locations. Beyond this example, it is possible to organize hemodynamics data differently. For example, if the columns represent hemodynamic data with varying different parameters (e.g., boundary conditions), then PCA can demonstrate the correlation among data with different parameters. Applying PCA to such data will enable data classification and pattern recognition, which could serve as a crucial step in machine learning. For example, linear discriminant analysis could be applied to labeled SVD/PCA modes for supervised learning and data classification~\cite{kutz2013data}. This technique has been used to investigate correlations between hemodynamic and geometric parameters and aortic insufficiency and ischemic events in the aorta of left ventricular assist device patients~\cite{Karmoniketal14}. PCA has  been used in predicting cerebral aneurysm rupture based on morphology and hemodynamics~\cite{Passerinietal12}. \edit{Statistical shape modeling with} PCA could facilitate geometric characterization of the \edit{vasculature~\cite{Mutetal14,Bruseetal16,Molonyetal19,Cosentinoetal20} and the heart~\cite{Hoeijmakersetal20,Fonsecaetal11}.} One challenge in using PCA is that the rows need to be consistently placed in different snapshots. For instance, if the hemodynamic data in columns are based on different patients, care must be taken to ensure that the data are oriented and aligned consistently. \edit{A major limitation of the PCA method is its inherent linearity. PCA cannot effectively identify low-dimensional nonlinear patterns in the data. The advantages gained by using manifold learning and nonlinear generalizations of PCA such as kernel PCA~\cite{scholkopf2018learning}, locally linear embedding~\cite{Ehlertetal19}, spectral clustering~\cite{vonLuxburg07}, and autoencoder neural networks~\cite{Agostini20} should be investigated in future work. 

 As with all statistical data analysis methods, sufficient quantity and quality of data is required such that the results obtained from applying PCA on a given dataset converges to the results that would be obtained given an infinite amount of data \cite{dauxois1982asymptotic} (unless one is only interested in the properties of the specific dataset itself).  This is dependent on a variety of factors, including the number of principal components to be identified, the amount of energy contained in each principal component, the accuracy required for any ensuing analysis, the presence of any bias in the sampled data towards certain principal components, and the presence of measurement noise that may be present in the data. Section \ref{sec:rpca} will discuss an extension of PCA that is explicitly designed to remove random noise from data.  In general, convergence of PCA can be evaluated by testing the extent to which adding or removing data influences the results, with convergence obtained if adding additional data does not have a tangible effect on the results of the analysis. 
}

\section{Robust principal component analysis (RPCA): noisy and fluctuating data}
\label{sec:rpca}
\subsection{Motivation} Noisy, fluctuating data can compromise the accuracy of PCA. Such data often result in a larger number of energetic PCA modes than what is truly needed to represent the physics of the data. The issue is pronounced in experimental data but can also exist in computational data. In experimental blood flow data, noisy and corrupt data are often inevitable. Additionally, in cases where transitional flow exists (e.g., blood flow in aneurysms~\cite{Valenetal11}), high-resolution CFD solvers can capture fluctuating velocity data that may compromise the temporal statistical correlation in PCA if one is interested in mean flow behavior. Robust principal component analysis (RPCA), as developed in \cite{candes2011robust}, and recently applied to fluids data in \cite{Scherletal20} has been proposed to overcome these limitations by separating the noisy data from the rest of the data using an optimization framework.

\subsection{Model theory and background}
\label{sec:RPCA}
 Consider the same data matrix $\mathbf{X}$ defined in Eq.~\ref{eqn:x}. We assume that the noise is randomly and sparsely distributed in the data. The first step is to write $\mathbf{X}$ as the sum of a low-rank matrix  $\mathbf{L}$ containing the correlated data we are interested in and a sparse matrix  $\mathbf{S}$ containing the noisy/fluctuating/corrupt data

 \begin{equation}
\mathbf{X} = \mathbf{L} + \mathbf{S} \;.
\end{equation}
Next, the above objective is mathematically formulated using an optimization problem
 \begin{equation}
 \label{eq:rpca1}
\min_ {\mathbf{L},\mathbf{S} } \left(\textup{rank}( \mathbf{L}) + \lVert \mathbf{S} \rVert_0\right)  \;  \textup{s.t.} \; \mathbf{X} = \mathbf{L} + \mathbf{S} \;,
\end{equation}
where $\textup{rank}( \mathbf{L})$ is equivalent to the number of nonzero singular values of $\mathbf{L}$, and $\lVert \mathbf{S} \rVert_0$ (which is not strictly a norm) denotes the number of nonzero entries in $\mathbf{S}$. Optimizing Eq.~\ref{eq:rpca1} is a combinatorial problem, which can typically only be solved using a non-tractable brute force method. Therefore, the problem is relaxed to a convex optimization problem, formulated as
 \begin{equation}
\min_ {\mathbf{L},\mathbf{S} } \left( \lVert \mathbf{L} \rVert_* + \lambda \lVert \mathbf{S} \rVert_{1} \right) \;  \textup{s.t.} \; \mathbf{X} = \mathbf{L} + \mathbf{S} \;,
\end{equation}
where $\lVert.\rVert_*$ is the nuclear norm (sum of the singular values), the nonconvex $l_{0}$ pseudonorm is replaced by the convex $l_{1}$ norm, and $\lambda$ is a regularization parameter controlling the level of  sparsity in $\mathbf{S}$. To find $\mathbf{L}$ and $\mathbf{S}$, the above problem can be solved using the augmented Lagrange multiplier algorithm~\cite{LinChenMa10}. Finally, we may use the $\mathbf{L}$ matrix to perform PCA as explained in the previous section. 

\subsection{Example: Low-dimensional behavior in hemodynamics data with noise }
\paragraph{\textit{Problem statement:}} Can RPCA improve low-dimensional data extraction in patient-specific abdominal aortic aneurysm (AAA) data with noise?

\begin{figure}[h!]
\centering
\includegraphics[scale=0.55]{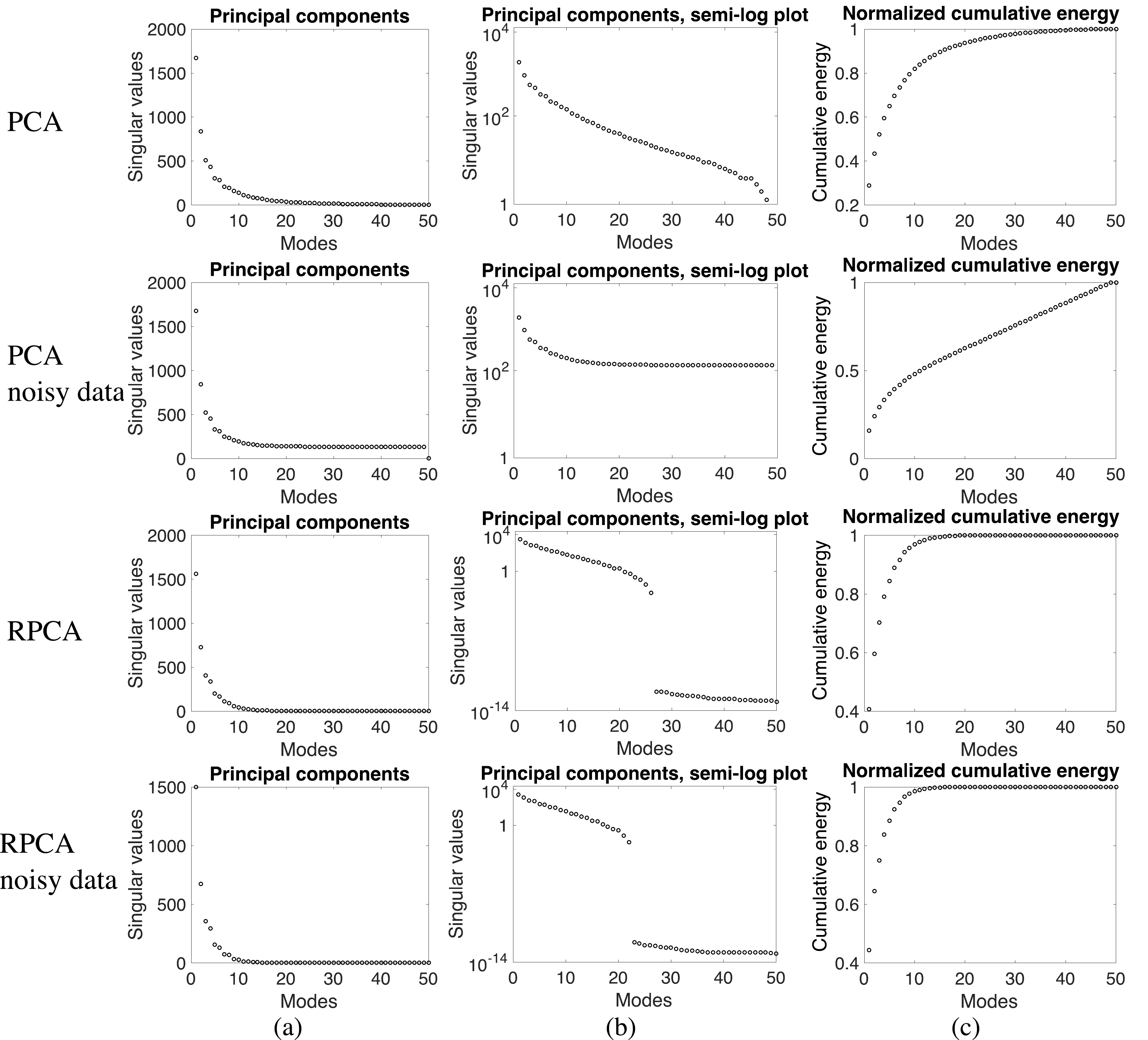}
\caption{Robust principal component analysis (RPCA) performed on abdominal aortic aneurysm velocity data. PCA on the original data (top row), PCA on the original data with noise (second row), RPCA on the original data (third row), and RPCA on the original data with noise (fourth row) are plotted where the principal components in (a) a regular plot, (b) semi-log plot, and (c) cumulative sum of singular values normalized by the sum of all singular values are shown.}
\label{fig:rpca}
\end{figure}

\paragraph{\textit{Problem solution:}}  The same AAA blood flow velocity data used in the previous section is used. Noise is randomly added in space and time in 30\% of the data. The value of noise is sampled from a random normal distribution with zero mean and a standard deviation equal to 10\% of the maximum streamwise velocity in the data. The noise is added to all components of the velocity vector using the same distribution. PCA and RPCA are applied to both original and noisy data. The results shown in Fig.~\ref{fig:rpca} demonstrate that PCA fails in exposing the low-dimensionality in noisy data where the singular values do not asymptote to zero. RPCA overcomes this issue and reveals the low-dimensionality in the data. An interesting observation is that RPCA applied to the original data (without noise) leads to a smaller number of dominant modes compared to PCA, due to the fact that some of the data is incorporated into the sparse matrix $\mathbf{S}$, and discarded prior to the decomposition. 


\subsection{Opportunities and challenges} RPCA is an excellent method to detect low-dimensionality and coherent patterns in experimental hemodynamics data where noise is inevitable. Recently, RPCA has been successfully applied to experimental particle image velocimetry (PIV) data of complex fluid flow~\cite{Scherletal20}. Experimental cardiovascular fluid mechanics techniques such as 4D flow magnetic resonance imaging (MRI) often carry a significant amount of noisy and corrupted data. RPCA could be applied to such data to reconstruct the dominant flow structures. Additionally, RPCA has the potential to facilitate data-driven modeling techniques such as dynamic mode decomposition (see Sec~\ref{sec:dmd}) that are sensitive to noise. One challenge in applying RPCA is the requirement that the corrupt data should be sparsely distributed. This requirement and sensitivity to the hyperparameter $\lambda$ should be investigated in the context of cardiovascular flows in future work. 


\section{Compressed sensing (CS): Reconstructing high-resolution data from low-resolution sampling} \label{sec:cs}
\subsection{Motivation} In experimental modeling of cardiovascular flows, high-resolution sampling of hemodynamics data is often not possible due to inherent limitations. However, high-resolution data are an essential part of many modeling tasks (e.g., patient-specific flow waveforms for inlet boundary conditions, spatial gradients for wall shear stress calculation, etc.). In the previous sections, we saw that high-resolution data may admit a low-dimensional sparse representation in an appropriate basis. The idea behind compressed sensing (CS) is that if we have a priori knowledge of such a basis, then we can reconstruct high-resolution data from low-resolution sampling. CS is a theoretical framework that solves an under-determined system of linear equations when the solution is known to be sparse. As a simple example, assume we have sampled velocity data in the aorta with a spatial resolution of 3mm (typical 4D flow MRI resolution), however, we would like to reconstruct the data with a spatial resolution of  0.5mm (typical CFD resolution). The only constraint here is that at the sampled MRI locations, the reconstructed data must match the 4D flow MRI data. Obviously, this is an under-determined system, meaning that the reconstruction problem has an infinite number of possible solutions. In CS, we regularize this problem to enable a solution, by promoting a solution that is sparse in the specified basis. 

\subsection{Model theory and background}  Suppose we have $m$ low-resolution measurements $\mathbf{y} \in \mathbb{R}^{m}$ and we want to reconstruct the high-resolution data $\mathbf{x} \in \mathbb{R}^{n}$ with $n$ components, where $n\gg m$. We know the location of low-resolution measurements with respect to the high-resolution data via a known measurement matrix $\mathbf{C}$
 \begin{equation}
\mathbf{y} = \mathbf{C} \mathbf{x} \;.
\label{eqn:ycx}
\end{equation}
If the low-resolution measurements are a subset of the high-resolution data, then each row of $\mathbf{C}$ consists of zeros everywhere except for the column corresponding to a measurement location, which has an entry of $1$.
The central assumption is that we have a priori knowledge of a transformation $\mathbf{\Phi}$ that relates the high-resolution data with a sparse vector $\mathbf{s}$ where $\mathbf{x} = \mathbf{\Phi} \mathbf{s} $, which implies that we know how to sparsify the data. Combining the previous two equations, we arrive at an equation that could be solved for $\mathbf{s}$
 \begin{equation}
\mathbf{y} = \mathbf{C}\mathbf{\Phi} \mathbf{s} \;.
\label{eqn:ycpx}
\end{equation}
This is an under-determined system of equations. To enable solution, the following optimization problem is solved

 \begin{equation}
\min_ {\mathbf{s}  }  \lVert  \mathbf{s} \rVert_1  \;  \textup{s.t.} \; \mathbf{y} = \mathbf{C}\mathbf{\Phi} \mathbf{s}  \;,
\label{eqn:mincs}
\end{equation}
where using the $l_1$ norm, we seek the sparsest solution $\mathbf{s}$ that satisfies the measurement constraints. Here, the use of the $l_{1}$ norm is again used as a convex proxy for the $l_{0}$ pseudonorm representing the number of nonzero elements of $\mathbf{s}$, in order to make the problem computationally tractable. Alternatively, the optimization problem could be written using the least absolute shrinkage and selection operator (LASSO) method~\cite{Tibshirani96}

 \begin{equation}
\min_ {\mathbf{s}  }  \lVert   \mathbf{y} - \mathbf{C}\mathbf{\Phi} \mathbf{s} \rVert_2  + \lambda   \lVert  \mathbf{s}  \rVert_1 \;,
\end{equation}
where the hyperparameter $\lambda$ controls the level of sparsity. Note in particular that this formulation means that Eq.~\ref{eqn:ycx} no longer needs to be exactly satisfied. Once $\mathbf{s}$ is found, the high-resolution data could be recovered using $\mathbf{x} = \mathbf{\Phi} \mathbf{s} $.

\subsection{Example: High-resolution reconstruction of blood flow waveform from low-resolution data}
\paragraph{\textit{Problem statement:}} Given low-resolution measured blood flow waveform data in the aorta, is it possible to reconstruct the high-resolution flow waveform using CS? 

\begin{figure}[h!]
\centering
\includegraphics[scale=0.55]{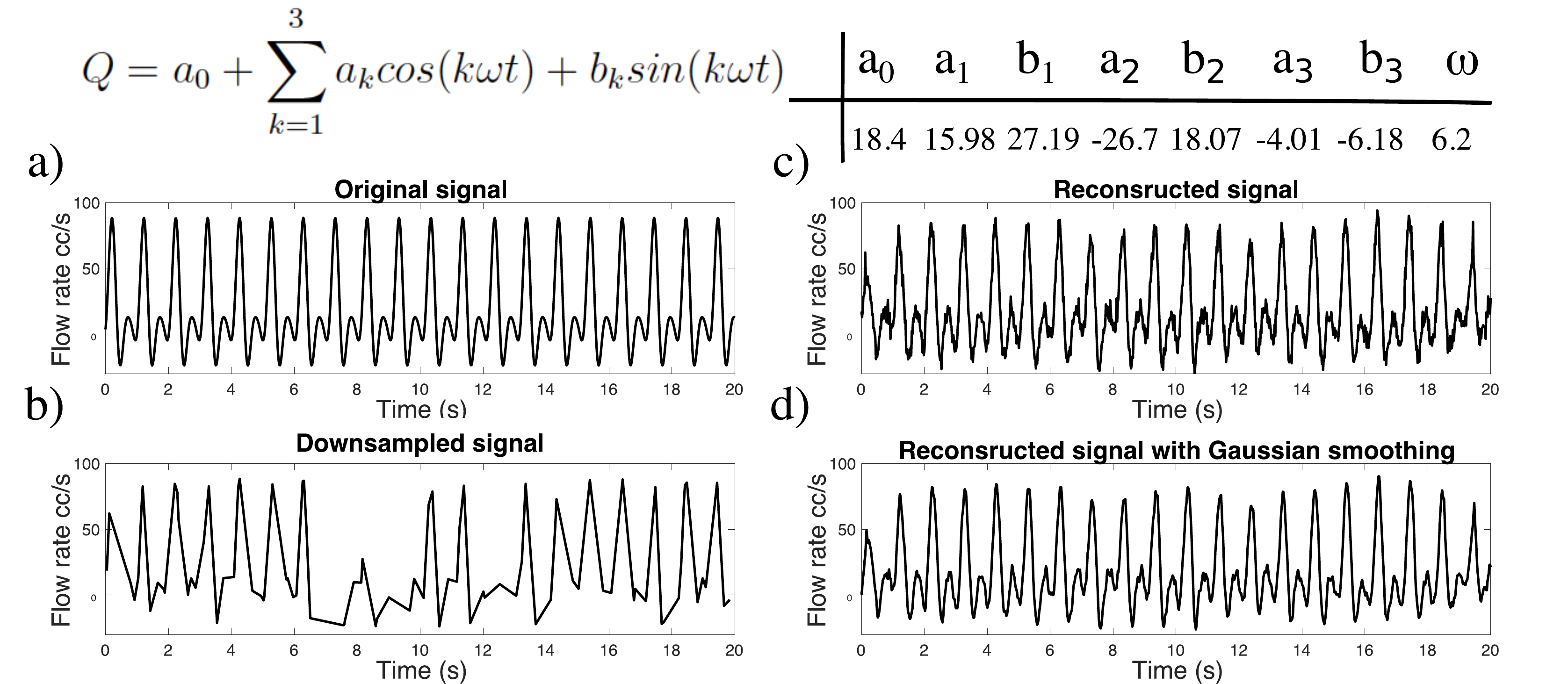}
\caption{Pulsatile blood flow rate waveform in the infrarenal aorta is presented over 20 cardiac cycles (T=1s) using Fourier series. (a) The original waveform, (b) the downsampled waveform, (c) compressed sensing waveform reconstruction, and (d) compressed sensing waveform reconstruction with Gaussian smoothing are plotted.     }
\label{fig:cs}
\end{figure}

\paragraph{\textit{Problem solution:}}  We consider a typical blood flow waveform in the infrarenal aorta represented using Fourier series (Fig.~\ref{fig:cs}). Such a waveform is often used as an inlet boundary condition for CFD simulations in abdominal aortic aneurysms~\cite{Arzani18}. In-vivo phase-contrast MRI (PCMRI) enables patient-specific measurement of such waveforms. We assume that we have low-resolution temporal sampling over 20 cardiac cycles. We randomly sample 100 data points over 20 cardiac cycles (on average 5 points per cardiac cycle). The goal is to reconstruct the data with 1000 data points (50 points per cardiac cycle). The original and downsampled waveforms are shown in Fig.~\ref{fig:cs}a and  Fig.~\ref{fig:cs}b, respectively. In this example, we assume that the waveform is sparse in the discrete cosine transform (DCT) basis. The CS solution (from Eq.~\ref{eqn:mincs}) is shown in Fig.~\ref{fig:cs}c. The result shows significant improvement over the low-resolution data. However, the CS waveform does not perfectly recover the original waveform even though the features are recovered. Artificial high-frequency features are present in the CS recovered waveform, which may be removed using a Gaussian smoothing filter (Fig.~\ref{fig:cs}d).

\subsection{Opportunities and challenges}
CS is a very important data-driven modeling tool with various applications~\cite{majumdar2018compressed}. In experimental cardiovascular biomechanics modeling, it is often not possible to collect data at the desired spatial and/or temporal resolution.  Under certain conditions, CS provides the means to recover high-resolution data given low-resolution measurements. If we know the basis where the data is sparse, then we do not necessarily need to collect high-resolution data as we can reconstruct the data using CS. In other words, if such a basis exists, then the Nyquist--Shannon sampling law~\cite{Shannon48}, which requires data sampling at a rate twice as fast as the highest frequency, is no longer required. CS is widely used in medical imaging such as computed tomography (CT)~\cite{ChenTangLeng08} and MRI~\cite{Jungetal09}. For example, CS has been used to accelerate blood flow velocity measurement in 4D flow MRI in the K-space (the Fourier transform of the magnetic resonance image)~\cite{Maetal19}. However, the application of CS to high-resolution reconstruction of low-resolution blood flow data has remained elusive. A major challenge is identifying the appropriate basis where the solution is sparse. In a number of applications, sparsity in space can be achieved via a wavelet transform, while temporal sparsity can often be obtained using a Fourier transform. For example, the wavelet transform has been used to successfully reconstruct spatial data governed by the diffusion equation with 10 times higher resolution~\cite{Ibanezetal19}. However, successful reconstruction using a standard basis will likely be challenging. Dictionary learning and transform learning provide the means to identify an optimized basis~\cite{majumdar2018compressed}, however, their success in complex blood flow data remains to be investigated. An alternative approach is to learn the sparsifying basis using training data. This topic is discussed in Sec.~\ref{sec:mlrom}. 

Another challenge is the requirement that the data should be sampled in a manner that is incoherent with the basis in which the signal is sparse.
This requirement, which can be expressed more formally as the restricted isometry property \cite{candes2008introduction}, ensures that the sampled signal can detect nonzero coefficients of all basis elements. As a counterexample, uniform sampling is ``periodic'' and thus is not incoherent with a Fourier basis. Most typically, we can allow this incoherence property to be satisfied by choosing a random sampling procedure.

\section{Data assimilation using the Kalman filter: Merging experimental and computational data}
\subsection{Motivation} Sometimes in cardiovascular flow modeling, we have the luxury of possessing both experimental data and computational models. However, both experimental and computational models have limitations and errors. Experimental data are often low-resolution and noisy, whereas even high-resolution computational simulations have limitations due to uncertainty in model parameters and/or governing equations. The field of ``data assimilation'' deals with hybrid models that integrate observed experimental data with computational models. The goal is to use the computational model to advance the solution in time and based on the availability of experimental data, alter the computational model's prediction to improve solution reliability.

\subsection{Model theory and background} 

Here, we will discuss the simplest implementation of the Kalman filter. Consider our computational model governed by a dynamical system as
\begin{equation} \label{eq:dynsysk}
\dot{\mathbf{x}}(t) = \mathbf{f}(\mathbf{x},t) + \mathbf{q}_1 \; ; \quad
{\mathbf{x}}(t_{0}) = \mathbf{x}_{0} + \mathbf{q}_2 \;,
\end{equation}
where $ \mathbf{q}_1$ is the unknown error in the model and  $\mathbf{q}_2$ represents the uncertainty in the initial condition of the system. Consider the set of experimental data as

\begin{equation}
\mathbf{y}(t) = \mathbf{g}(t) + \mathbf{q}_3 \;,
\end{equation}
where $\mathbf{g}(t) $ are the measured data and  $\mathbf{q}_3$ is the associated error. Note that the measurements do not necessarily need to be available at all points.  We would like our model (Eq.~\ref{eq:dynsysk}) to satisfy the experimental observation. The above equations form a system of overdetermined equations where we have more equations than unknowns. Therefore, an optimization problem is defined to find the solution that minimizes the error weighted by the above errors in the model and experimental data. In the case of one-dimensional data, the solution to such optimization problem may be written as

\begin{equation} \label{eqn:K}
\bar{x}_{k} = x_{k} + K ( y_k - x_k) \;,
\end{equation}
where $\bar{x}$ is the data assimilated prediction, $x$ is the computational/analytical model, $y$ denotes the experimental data, $K$ is the Kalman filter (Kalman gain), and the subscript $k$ specifies the time-step. The experimental data may not be available at every model time-step, therefore the above prediction will only be applied at the time-steps where the experimental data are available. The Kalman gain may be written as

\begin{equation}
K = \frac{ \sigma^2_x   }{  \sigma^2_x  +  \sigma^2_y    } \;,
\end{equation}
where it is assumed that the errors could be represented with Gaussian distributions. $\sigma_x$ and $\sigma_y$ represent the standard deviation of the error distribution in the model and experimental data, respectively. It could be seen that $K$ weights the contribution of experimental and computational data based on their expected errors. 

The above implementation could be extended to vectorial data and the Kalman gain could be improved and updated at every step using the extended Kalman filter method. These methods are described in a simple but informative presentation in~\cite{kutz2013data} and in more detail in~\cite{evensen2009data,asch2016data}.

\subsection{Example: Merging computational and experimental trajectories in transient vortical flows}\label{sec:hill}
\paragraph{\textit{Problem statement:}} Consider an unsteady version of Hill's spherical vortex, which is an extension to the steady version that has been used as a very simple analytical model of 3D vortex flows in the left ventricle of the heart~\cite{FalahatpishehPedrizzettiKheradvar14}. The goal is to construct particle trajectories considering uncertainty in the initial condition and availability of noisy experimental data.  

\paragraph{\textit{Problem solution:}} We extend Hill's spherical vortex~\cite{FalahatpishehKheradvar15} to a transient vortex by periodic movement of the vortex center:

\begin{equation} \label{eqn:hill}
\mathbf{u}(x,y,z,t) =
\begin{cases}
 (x^2 + 1 - 2r^2, \;  \;xy, \; \; xz)^T , & \mbox{if } r\leq 1 \\ (z^2r^{-5} - \frac{1}{3}r^{-3} - \frac{2}{3}, \;  \;xyr^{-5}, \; \; xzr^{-5})^T , & \mbox{if } r>1 
\end{cases}
\end{equation}
where $r=\sqrt{ x^2 + (y-A)^2 + (z-A)^2}$ and $A = \sin (\pi t/50) $. The above equation is numerically integrated for T=20, with the initial condition $\mathbf{x}_0 = (0, 0.5, 0)^T$ and time-step $\Delta t$ = 0.001 to generate the reference data. We assume synthetic experimental data is available at a lower sampling rate ($\Delta t$ = 0.04) with uncorrelated Gaussian white noise noise added with $\mathbf{\sigma}_y$=0.05. For the computational model, we assume we do not have precise knowledge of the initial condition, and thus to each component of  $\mathbf{x}_0$ we add Gaussian noise with $\mathbf{\sigma}_x=0.15$. The Kalman filter method is used to correct the model prediction based on the synthetic experimental data using Eq.~\ref{eqn:K} for each component of $\mathbf{x}$, which is applied once every 40 steps (based on the lower sampling rate of the experimental data). 

\begin{figure}[h!]
\centering
\includegraphics[scale=0.55]{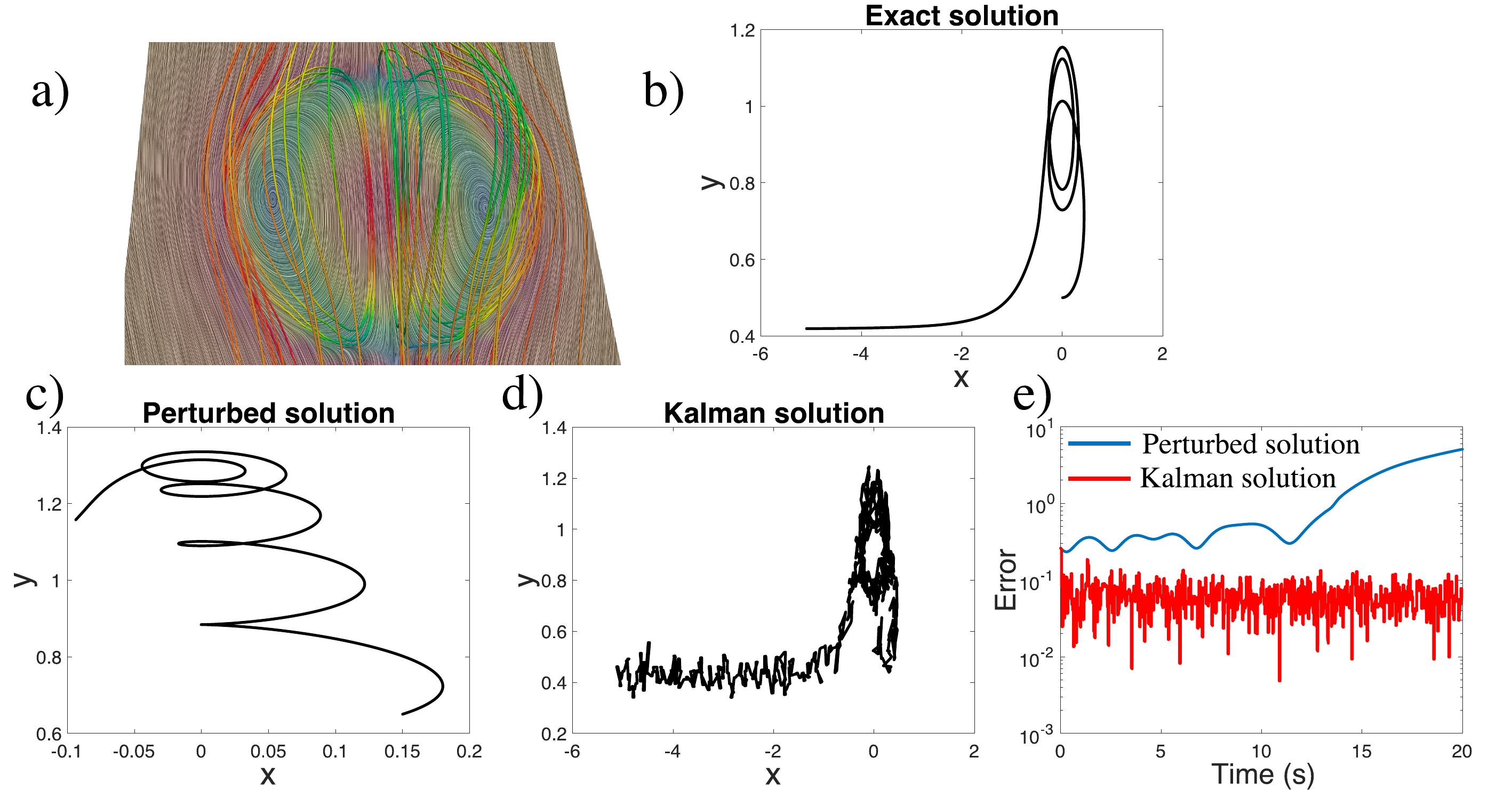}
\caption{Data assimilation based on Kalman filter is applied to a sample trajectory from the transient version of Hill's spherical vortex. (a) 3D streamlines and 2D streamlines in a cross-section are shown at the initial time-step.  The trajectory of the (b) exact solution, (c) perturbed solution (uncertain initial condition), (d) Kalman filter solution are shown in the 2D x-y plan.  (e) The errors in the perturbed and Kalman filter solutions are plotted (semi-log plot).   }
\label{fig:kalman}
\end{figure}

The results are shown in Fig.~\ref{fig:kalman}. A snapshot of the vortex flow is shown. The induced unsteadiness in the flow causes trajectory sensitivity to the initial condition. We can see that the perturbed solution (Fig.~\ref{fig:kalman}c) behaves very differently from the exact solution  (Fig.~\ref{fig:kalman}b). The Kalman filter improves the estimation of the trajectory (Fig.~\ref{fig:kalman}d), however, noise is still present due to the noisy nature of the experimental observation. Finally, the errors in trajectories are shown (Fig.~\ref{fig:kalman}e) in a semi-log plot where the Kalman filter significantly outperforms the perturbed solution.

\subsection{Opportunities and challenges}
Herein, we have presented a very simple example of the Kalman filter. More advanced implementations of the Kalman filter are possible for more reliable results in more complex settings. For example, the ensemble Kalman filter has been used to merge 4D flow MRI and CFD simulations~\cite{Gaidziketal19}. Parameters in 0D and 1D blood flow models have been estimated based on clinical measurements using the ensemble Kalman filter~\cite{Devaultetal08,Canutoetal20} and the unscented Kalman filter~\cite{Pantetal14}, which are extensions formulated to deal with large state dimensions and nonlinear system dynamics, respectively. The ensemble Kalman filter has also been used to estimate the inlet flow waveform in patient-specific arterial network models~\cite{Arnoldetal17}. Reduced-order unscented Kalman filter has been used in conjunction with fluid-structure interaction (FSI) simulations to estimate aortic aneurysm wall stiffness from wall displacement measurements~\cite{Bertoglioetal12}. Other data assimilation methods such as variational data assimilation could also be used to merge multi-modality hemodynamics data~\cite{Funkeetal19}. In general, the family of Kalman filter data assimilation models provides the means to improve the accuracy of computational models by leveraging even imperfect experimental data. There are a growing number of studies in the literature on merging CFD and 4D flow MRI data using data \edit{assimilation~\cite{Rispolietal15,Fathietal18,KoltukluougluBlanco18,Gaidziketal19,Annioetal19,Ferdianetal20}.} PIV is another popular approach in hemodynamics quantification. Multi-modality data assimilation methods that leverage all of these modalities could improve the accuracy and reliability needed in transformative patient-specific hemodynamics modeling. \edit{It would be interesting to compare the relative success of all of these methods in future work.} One challenge in Kalman filter modeling is the prior knowledge of the error distribution and magnitude, which might not be always available and needs to be estimated. Additionally, the assumption that errors follow a Gaussian distribution might not always be appropriate. 

\section{Reduced-order physics: Dynamic mode decomposition (DMD) and proper orthogonal decomposition (POD)} \label{sec:dmd}
\subsection{Motivation} Discovering low-dimensional behavior in blood flow data enables a simplified understanding of the flow physics and opens up the door for efficient reduced-order and data-driven modeling. Physically, these reduced-order models are motivated by the presence of coherent structures or patterns, which are often observed in even highly-complicated fluid flow processes.  Proper orthogonal decomposition (POD) has been traditionally used in fluid mechanics for reduced-order modeling of coherent structures~\cite{lumley2007stochastic,BerkoozHolmesLumley93,Holmesetal12}. POD guarantees optimal reconstruction of the original dataset from an energetic perspective for a given number of modes. However, this can come with the price of losing physical interpretability due to the associated frequency mixing~\cite{Noack16}.  Dynamic mode decomposition (DMD) is a more recent technique, which is  data-driven (equation-free) and reduces temporal snapshots of input data to a best-fit linear reduced-order model~\cite{Schmid10,Rowleyetal09,Tuetal13,Kutzetal16a}. POD ranks modes by energy (reconstruction optimality), whereas DMD decomposes data into modes that isolate different dynamical features within a dataset.

\subsection{Model theory and background}
\label{sec:modalTheory}

Generally speaking, modal decomposition methods seek to decompose a system, or dataset, into spatial functions (modes), and their temporal coefficients.  That is, if we have a system state given by $\mathbf{y}(x,t),$ we wish to obtain a decomposition that separates the spatial and temporal features by
\begin{equation}
\label{eq:modaldecomp}
\mathbf{y}(x,t) \approx \sum_{j=1}^m  a_j(t) \mathbf{u}_j(x) \;,
\end{equation}
where $\mathbf{u}_j(x)$ are spatial functions, with time-varying coefficients given by $a_{j}(t)$. 
There are several methods to obtain such a decomposition from data, each of which has certain desirable features. If we wish to obtain a decomposition that is most efficient in the sense that the difference between the left and right sides of equation~\ref{eq:modaldecomp} is minimized, then we can utilize a singular value decomposition (i.e. PCA), as described in section \ref{sec:PCA}, utilizing the optimality property of SVD that minimizes the reconstruction error (Eq.~\ref{eq:SVDerror}). Here, the spatial modes $\mathbf{u}_j$ correspond to the left singular vectors of the data matrix, with the temporal functions given by the scaled right singular vectors, $a_j(t) = \sigma_{j} \mathbf{v}_j^{*}$. In this context, this is called a proper orthogonal decomposition. As the name suggests, this decomposition also gives spatial modes that are orthogonal (from the properties of SVD). 

In practice, if the size of a snapshot is much larger than the number of snapshots (i.e., $n \gg m$), it can be computationally cheaper to compute this decomposition by first finding $\mathbf{v}_j$ from the eigendecomposition given in Eq.~\ref{eq:PODsnapshots}, from which the POD modes can be subsequently found from 
\begin{equation}
\mathbf{u}_j = \sigma_{j}^{-1}\mathbf{\tilde X} \mathbf{v}_j \;.
\end{equation}
This method for computing POD was first proposed in \cite{Sirovich-87}, and is the approach employed in the examples considered here. 
 \edit{As discussed previously in the context of PCA, sufficient data is required to ensure that the identified POD modes are sufficiently converged. As an example, in the context of noisy experimental data, 
 \cite{epps2010error} suggests a threshold of $\sigma_j > \epsilon\sqrt{mn}$ to ensure the accuracy of the first $J$ modes, where $\epsilon$ is the root-mean-square noise level, and as before $m$ and $n$ denote the number of snapshots and size of each snapshot of data, respectively.  Note also that the accuracy requirements also depend on how the results of POD are used.  For example, higher quantitative accuracy may be required if the modes are used for reduced-order modeling via Galerkin projection \cite{Holmesetal12}, as opposed for qualitative identification of coherent structures.}

POD is not the only choice of decomposition for Eq.~\ref{eq:modaldecomp}. We might also want to obtain a decomposition where the coefficients $a_{j}(t)$ possess certain properties (note that they are also orthogonal functions in the POD). In particular, if we have 
\begin{equation}
a_{j}(t) = \exp(\lambda_{j}t),
\end{equation}
then each spatial mode can be associated with a characteristic growth/decay rate and frequency of oscillation in time, given by the real and imaginary components of $\lambda_{j}$, respectively.  Such a decomposition can be obtained via dynamic mode decomposition (DMD). For DMD, it is assumed that the data is time-resolved, with a fixed timestep, $\Delta t$. The fundamental idea underlying DMD is to find a linear operator $\mathbf{A}$, which maps the system one timestep into the future, the eigenvectors and eigenvalues of which are DMD modes and eigenvalues. 
Letting $\mathbf{ X}_{1}$ and $\mathbf{ X}_{2}$ denote the data matrix with the last and first column (snapshot) removed respectively, the linear operator $\mathbf{A}$ is given by 
\begin{equation}
\mathbf{A} = \mathbf{ X}_{2} \mathbf{ X}_{1}^{+} \;,
\end{equation}
where $^{+}$ denotes the pseudoinverse.  In practice, the $n\times n$ matrix $\mathbf{A} $ is typically not computed directly, as it is computationally easier to work in a lower dimensional space obtained via an SVD of the data. A typical algorithm to compute DMD involves the following steps:
\begin{enumerate}
\item Compute the (optionally truncated to rank $r$) SVD  $\mathbf{X}_1 = \mathbf{U}\mathbf{\Sigma}\mathbf{V}^* \approx \mathbf{U}_{r}\mathbf{\Sigma}_{r}\mathbf{V}_{r}^*$. 
\item Compute $\mathbf{A}_{r} =  \mathbf{U}_r^* \mathbf{X}_{2} \mathbf{V}_r\mathbf{\Sigma}_r^{-1}$, and find it's eigenvalues and eigenvectors satisfying $\mathbf{A}_{r}\mathbf{w}_j = \mu_j \mathbf{w}_j$
\item Compute DMD modes $\mathbf{\phi}_{j} = \mathbf{U}_{r}\mathbf{w}_j $ corresponding to each continuous time eigenvalue $\lambda_{j}$, where $\mu_{j} = \exp(\lambda_{j}\Delta t)$.
\end{enumerate}
With this approach, the number of DMD modes (or equivalently, the number of terms in the sum in Eq.~\ref{eq:modaldecomp}) is controlled by the truncation parameter, $r$. In practice, alternative approaches to choose a smaller number of modes can include modifying the basis of projection (e.g.,~\cite{Wynnetal13,sashittal2019reduced}), choosing a smaller subset of modes to use from those that are computed (e.g.,~\cite{JovanovicSchmidNichols14,drmac2018data,zhang2020evaluating}), or performing a balanced truncation on the resulting linear model (e.g.,~\cite{RowleyDawson17}). 
There are numerous further variants of DMD that extend its applicability and usefulness for a wider variety of systems, incorporating, for example, external inputs \cite{Proctoretal16}, nonlinear observables \cite{Williamsetal15}, and noisy data \cite{Dawsonetal16,hemati2017tls,askham2018variable}. For a more comprehensive discussion of these variants, see~\cite{RowleyDawson17,Tairaetal17,brunton2019data}. 



\subsection{Example: Blood flow physics in a cerebral aneurysm using DMD and POD} \label{sec:dmde}
\paragraph{\textit{Problem statement:}} Given spatiotemporally resolved velocity data in a 2D cerebral aneurysm model, compare the dominant DMD and POD modes. 

\begin{figure}[h!]
\centering
\includegraphics[scale=0.45]{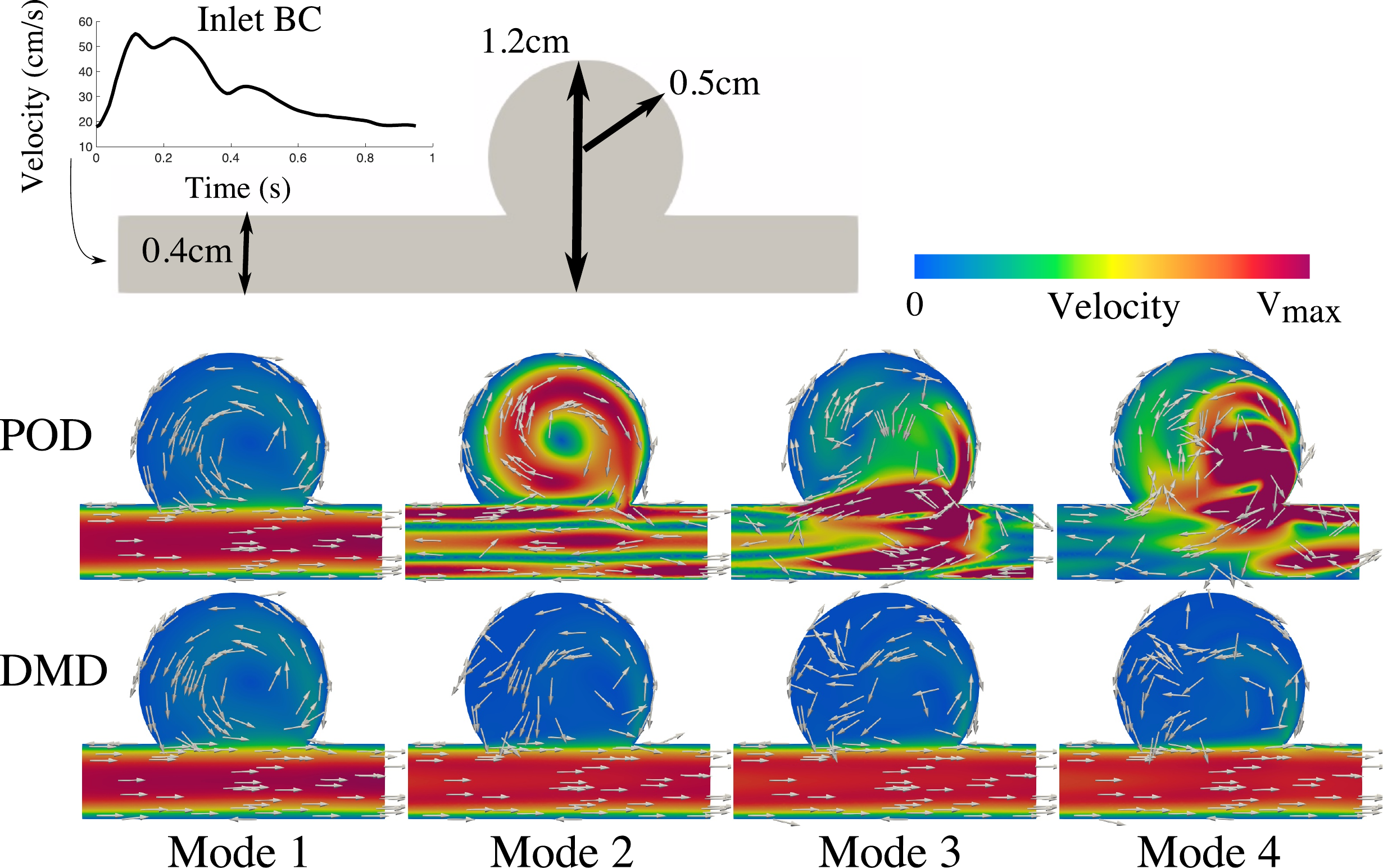}
\caption{The first four proper orthogonal decomposition (POD) and dynamic mode decomposition (DMD) modes are shown in an idealized aneurysm model with the shown geometric dimensions. The pulsatile waveform used as inlet boundary condition is shown. }
\label{fig:dmd}
\end{figure}

\paragraph{\textit{Problem solution:}} Transient CFD simulations are performed using the open-source finite-element solver FEniCS~\cite{LoggMardalWells12}. An idealized 2D cerebral aneurysm model is considered with the geometric dimensions shown in Fig.~\ref{fig:dmd}. Incompressible Navier-Stokes equations are solved with $\mu$=0.04 P and $\rho$=1.06 g/cm$^3$. The pulsatile inlet flow waveform is shown in Fig.~\ref{fig:dmd} and the simulations are run for three cardiac cycles and the last cycle is used in data analysis. The mesh consists of 15.2K triangular elements. The first four dominant POD and DMD modes are shown in Fig.~\ref{fig:dmd}. The dominant DMD modes are selected using the sparsity-promoting algorithm given in~\cite{JovanovicSchmidNichols14}.  The first dominant mode in both methods is very similar. POD picks up more complex structures in the other modes, whereas the DMD modes seem to be influenced by the dominant vortex in the aneurysm. 



\subsection{Opportunities and challenges} Modal analysis techniques such as POD and DMD facilitate flow physics analysis by breaking down the flow evolution into  several coherent structures (modes) ranked based on their energy (POD) and dynamics (DMD). These reduced-order models could also be used to reduce computational costs in simulations. However, accurate long-term extrapolation beyond the training data used in deriving these models remains a challenge. POD has been used in characterizing turbulent blood flow~\cite{Grinbergetal09,KefayatiPoepping13} and DMD has been used for quantitative modal analysis of blood flow physics~\cite{DiKadem19,HabibiDawsonArzani20}. The pulsatile nature of blood flow and variability in flow patterns during different phases of the cardiac cycle challenge the direct application of these  techniques.  Multistage DMD with control has been proposed to overcome these challenges in blood flow physics analysis with DMD~\cite{HabibiDawsonArzani20}.

\section{Machine learning reduced-order models (ROM): Overcoming uncertainty in computational models using low-fidelity experimental data}\label{sec:mlrom}
\subsection{Motivation} Uncertainty in model parameters is often an inevitable aspect of patient-specific computational hemodynamics simulations. Inlet flow waveform, flow split ratio among the outlets, blood rheology, the 3D geometry, and vessel wall material property in FSI simulations are often not precisely known. These uncertainties may question the fidelity of even high-resolution and minimally dissipative CFD solvers, once the results are being interpreted in the context of personalized medicine. On the other hand, in-vivo flow measurement techniques such as 4D flow MRI do not suffer from such limitations, however, they are limited in resolution and produce noisy data. Machine learning reduced-order models provide a framework where one can construct a library of high-resolution computational simulations by varying the uncertain parameters, and couple that with reduced-order modeling and compressed sensing to reconstruct high-resolution data and identify the uncertain parameter using the low-resolution experimental data~\cite{BrightLinKutz13}. 

\subsection{Model theory and background} First, we perform several high-resolution computational simulations by varying the uncertain parameter $p$ times ($\beta_1, \beta_2, \cdots, \beta_p$). Each simulation often results in several data snapshots in time. POD (equivalently, PCA) is performed on the dataset created for each parameter. Assume we need to keep $m$ POD modes for each simulation such that we can reconstruct the data with acceptable error (e.g., capturing 99\% of the energy). The number $m$ could be different for each set of simulations, depending on the change in data complexity with variation in the uncertain parameter.  A library could be constructed as

 \begin{equation}
\mathbf{\psi} = \begin{bmatrix} \mathbf{\phi}_{1}(\mathbf{x},\beta_1)  & \cdots &  \mathbf{\phi}_{m}(\mathbf{x},\beta_1) &   \mathbf{\phi}_{1}(\mathbf{x},\beta_2)  & \cdots &  \mathbf{\phi}_{m}(\mathbf{x},\beta_2) & \cdots &  \mathbf{\phi}_{1}(\mathbf{x},\beta_p)  & \cdots &  \mathbf{\phi}_{m}(\mathbf{x},\beta_p)        \end{bmatrix} \;.
\end{equation}

This library contains a reduced-order representation of all possible dynamics in the system within the range of the considered parameters and may be thought of as a supervised machine learning step. Note that it is possible to perform POD just once on the entire appended dataset as opposed to each $\beta$ parameter. This approach could still be used for data reconstruction, however, it will make parameter identification less trivial. Next, an underdetermined system of equations is constructed as $\hat{\mathbf{Y}} = (\mathbf{C}\mathbf{\psi} ) \mathbf{b} $, where $\hat{\mathbf{Y}}$ is the low-resolution experimental data, $\mathbf{C}$ is a measurement matrix specifying measurement locations, and $\mathbf{b}$ is a vector to be determined. To find the solution, the following compressed sensing problem is solved (see Sec.~\ref{sec:cs})

 \begin{equation}
\min_ {\mathbf{b}  }  \lVert  \mathbf{b} \rVert_1  \;  \textup{s.t.} \; \hat{\mathbf{Y}} = (\mathbf{C}\mathbf{\psi} ) \mathbf{b} \;.
\label{eqn:mlcs}
\end{equation}
To enable compressed sensing, the measurement locations used ($\mathbf{C}$) should be randomly selected. As we discuss in Example 2 below, relaxing the compressed sensing optimization problem can perform better in parameter identification

 \begin{equation}
\min_ {\mathbf{b}  }  \lVert  \mathbf{b} \rVert_1  \;  \textup{s.t.} \;   \lVert  \hat{\mathbf{Y}} - (\mathbf{C}\mathbf{\psi} ) \mathbf{b} \rVert_2  < \epsilon \;,
\label{eqn:mlcs2}
\end{equation}
where $\epsilon$ is a tolerance.  Once the above optimization problem is solved, the high-resolution data is reconstructed as  $\mathbf{Y} = \mathbf{\psi} \mathbf{b}$. 

Thus far, we have not specified the locations of measurements (as defined by $\mathbf{C}$).  Additionally, we have assumed knowledge of a basis in which our data is sparse, but have not prespecified which basis elements will have nonzero coefficients (i.e., which terms in $\mathbf{b}$ are nonzero). Here, following the work of \cite{manohar2018sensor}, we show that favorable sensor/measurement locations can readily be found given knowledge of such basis elements. From section \ref{sec:PCA}, a low-dimensional set of functions that capture most of the original data may be obtained from the columns of $\mathbf{U}_r$ in the truncated SVD
\begin{equation}
\mathbf{\psi} = \mathbf{U} \mathbf{\Sigma}\mathbf{V}^* \approx  \mathbf{U}_r \mathbf{\Sigma}_r\mathbf{V}_r^*,
\end{equation}
where $\mathbf{U}_r$ and $\mathbf{V}_r$ denotes the first $r$ columns of $\mathbf{U}$ and $\mathbf{V}$, and $\mathbf{\Sigma}_r$ contains the first $r$ diagonal entries of $\mathbf{\Sigma}$. Under the assumption that $\mathbf{\psi} \approx \mathbf{U}_r\mathbf{U}_r^* \mathbf{\psi}$, we can, in theory, reconstruct the high-resolution data from  $r$ measurement locations by solving the (no longer underdetermined) equation
\begin{equation}
\label{eq:robustSensors}
\hat{\mathbf{Y}} = \mathbf{C} \mathbf{U}_r ( \mathbf{U}_r^*  \mathbf{\psi} )  =  \mathbf{C} \mathbf{U}_r \mathbf{a},
\end{equation}
where the $r$-dimensional vector of coefficients $\mathbf{a} =  \mathbf{U}_r^*  \mathbf{\psi} $.
In principle, equation \ref{eq:robustSensors} can be solved for $\mathbf{a}$ for any choice of sensor locations $\mathbf{C}$ that give linearly independent equations, by using the pseudoinverse
\begin{equation}
\label{eq:Finda}
\mathbf{a} = \left( \mathbf{C} \mathbf{U}_r \right)^+ \hat{\mathbf{Y}}.
\end{equation}
However, some choices of $\mathbf{C}$ can result in these equations being ill-conditioned, which means in particular that noisy measurements $ \hat{\mathbf{Y}}$ can give wildly inaccurate reconstructions of the full data. One method to select measurement locations, as described in \cite{manohar2018sensor} is to choose sensor locations in a manner that results in $\mathbf{C} \mathbf{U}_r$ having a small condition number. This can be achieved via the pivoted QR decomposition
\begin{equation}
\label{eq:QR}
\mathbf{U}_r^T\mathbf{C}^T = \mathbf{Q}\mathbf{R}, 
\end{equation}
where the permutations in $\mathbf{C}^T$ ensure that the diagonal elements of $\mathbf{R}$ are ordered from largest to smallest. In the examples below, we call this approach optimal sampling.  Similar data reconstruction methods are used in the empirical interpolation methods described in \cite{barrault2004empirical,chaturantabut2010deim,drmac2016new}.
 To summarize, application of this method involves
 \begin{enumerate}
 \item Obtain a set of modes/functions $\mathbf{U}_r$ that sufficiently capture the high fidelity data (e.g.~via an SVD).
 \item Find optimal measurement locations (e.g.~low-dimensional data) using a pivoted QR decoposition of $\mathbf{U}_r$ (equation \ref{eq:QR}).
 \item Find the mode coefficients $\mathbf{a}$ from equation~\ref{eq:Finda}.
 \item Reconstruct the high-fidelity data via $\mathbf{Y} = \mathbf{U}_r \mathbf{a}$.
  \end{enumerate}

\subsection{Example 1: Reconstruct high-resolution Womersley flow with uncertain Womersley number and optimal sensor placement }
\paragraph{\textit{Problem statement:}} Given a few clean or noisy velocity samples from the Womersley flow profile, is it possible to reconstruct the entire velocity profile without knowing the Womersley number? 

\begin{figure}[h!]
\centering
\includegraphics[scale=0.45]{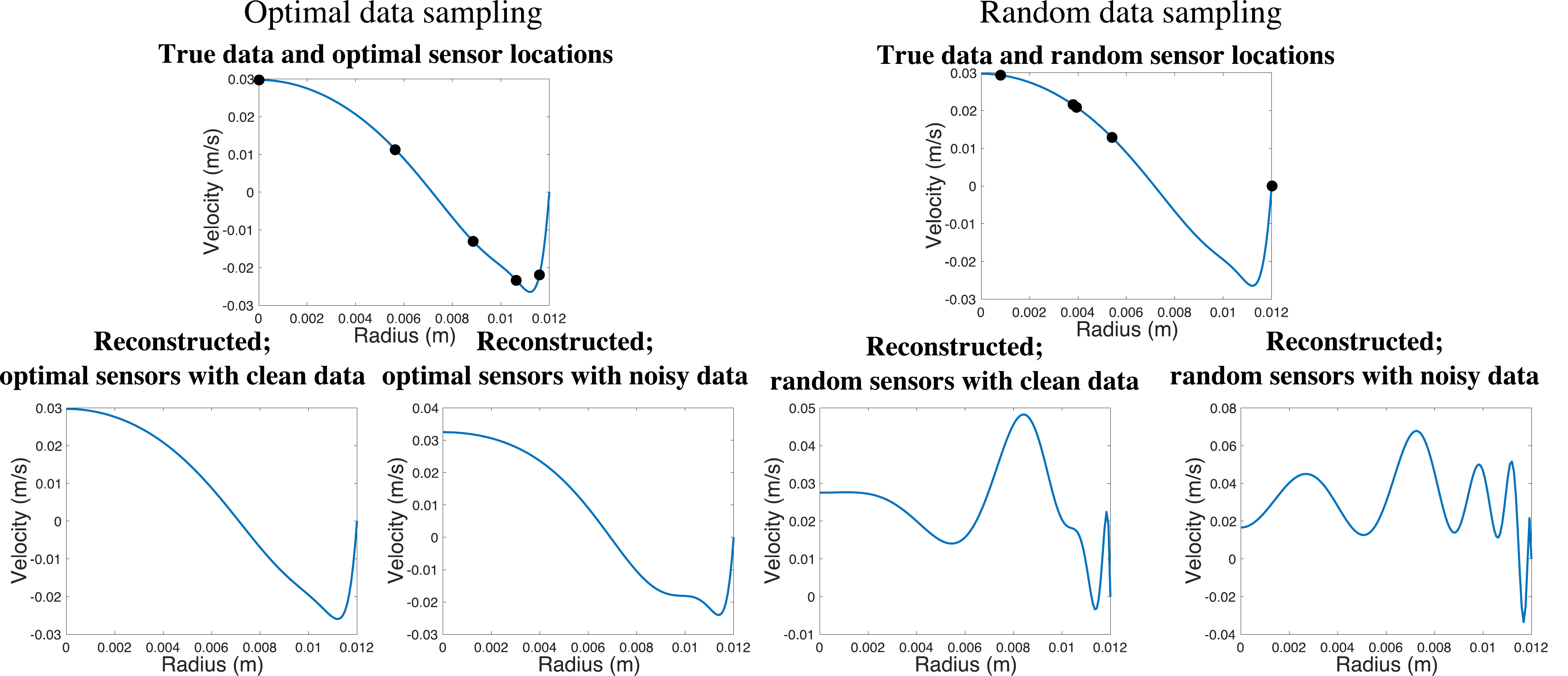}
\caption{The groundtruth velocity profile is shown with optimal (left) and random (right) sampling. For each case, the reconstructed velocity profile under clean and noisy data is shown.  }
\label{fig:mlos}
\end{figure}

\paragraph{\textit{Problem solution:}}  For demonstration purposes, we consider a wide range of Womersley numbers $\alpha$ (1--29) based on a fixed pressure gradient waveform (reported in~\cite{HabibiDawsonArzani20}) to construct a library of Womersley velocity profiles  based on Womersley's analytical solution~\cite{Womersley55}. The data is based on 150 radial locations and 100 time-steps over one cardiac cycle. In this example, in constructing the library, we apply POD once to the entire data.  The reference data (groundtruth) is considered to be $\alpha$=14.7. To test how the method works under noise, noisy data is created by adding random Gaussian noise with zero mean and standard deviation of 10\% the maximum velocity. Given only 5 random and optimal sample points along the radial direction at one time-step, we are interested in reconstructing the  velocity profile. The results are shown in Fig.~\ref{fig:mlos}. Optimal sensor placement can reconstruct the true data with high accuracy ($l_{2}$ norm in error 0.03 and 0.002 for noisy and clean data, respectively). Data reconstruction with random sensor placement has difficulty in reconstructing the data ($l_{2}$ norm in error 0.45 and 0.35 for noisy and clean data, respectively). More sampling points are required for successful reconstruction with random sampling, and the optimal sensor placement method is more robust to noise. 

\subsection{Example 2: Reconstruct high-resolution cerebral aneurysm flow with uncertain viscosity }\label{sec:csEx2}
\paragraph{\textit{Problem statement:}} Patient-specific variation in blood viscosity is common in patients with hemorheological disorders (e.g., diabetes)~\cite{ChoMooneyCho08}. Given sparse (low-resolution) sampling of noisy velocity data in a 2D cerebral aneurysm flow with an unknown viscosity, is it possible to identify the viscosity? 

\begin{figure}[h!]
\centering
\includegraphics[scale=0.6]{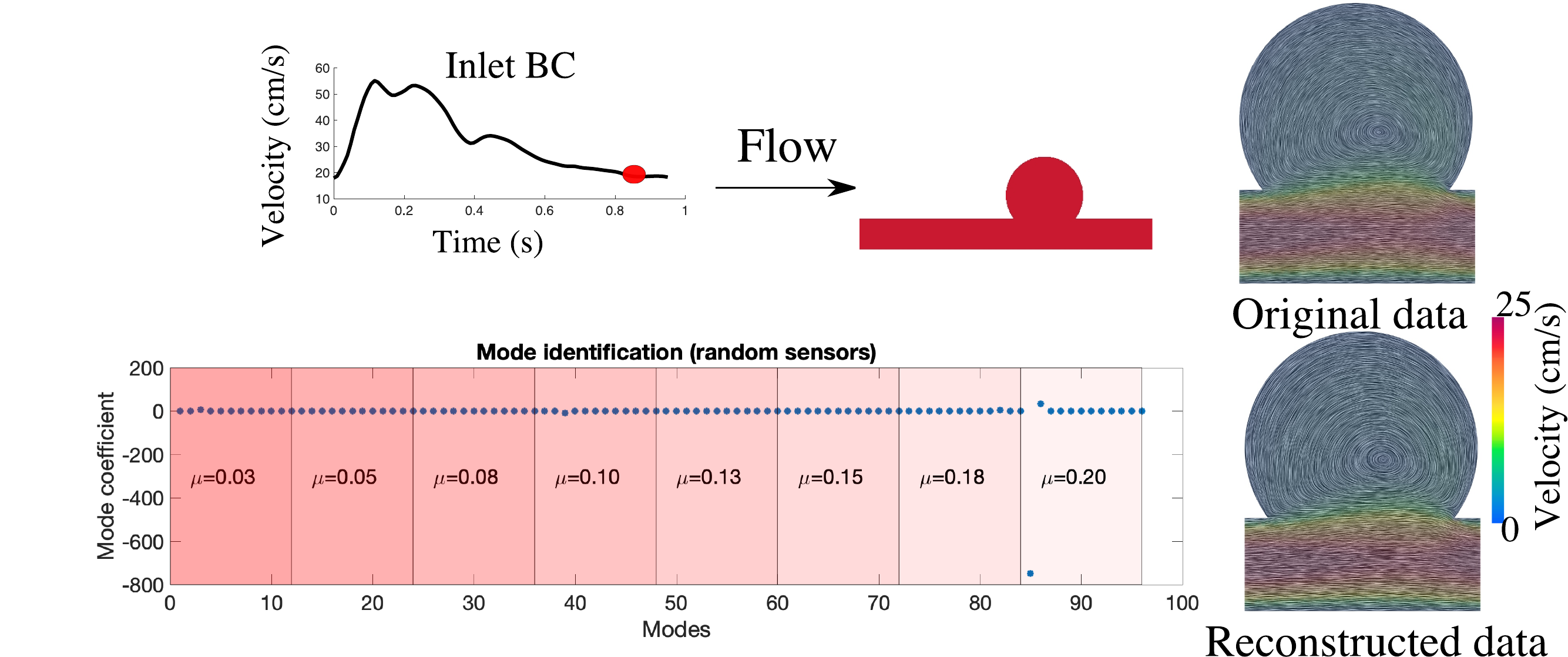}
\caption{The 2D aneurysm model and inlet boundary condition for CFD simulations are shown. The velocity streamlines for the original and reconstructed data are shown. The time-point of the original data is marked with a red dot in the waveform. The mode coefficients in the reduced-order model library are plotted. The method correctly identifies the dominant mode as being associated with the set of $\mu$=0.2 P modes. The measurement sensors were randomly placed.}
\label{fig:mlrom}
\end{figure}

\paragraph{\textit{Problem solution:}} The 2D cerebral aneurysm model in Sec.~\ref{sec:dmd} is considered. Simulations are performed for 8 different dynamic viscosity values uniformly distributed between $\mu$=0.03--0.2 P. Spatiotemporal velocity data are collected for all $\mu$ values. We consider $\mu$=0.2 P as the groundtruth data. Additionally, we assume we only have one temporal snapshot inside the cardiac cycle available for measured data. In this example, we consider 150 random data samples (sensors) to collect 2D velocity vector data.  Noisy data are created by adding Gaussian noise with zero mean and standard deviation of 5\% the velocity data at each location. POD is performed for each parameter and appended to build the library $\mathbf{\psi}$ by keeping 12 modes for each parameter. The relaxed version of the compressed sensing problem (Eq.~\ref{eqn:mlcs2}) is solved to identify the active modes in the library and reconstruct the data. Parameter identification is done based on the dominant active modes in the library.  The resulting active modes are shown in Fig.~\ref{fig:mlrom}. The dominant active mode belongs to the set of modes where $\mu$=0.2 P, and therefore the method correctly identifies the unknown viscosity. Most of the other modes are close to zero in accordance with the sparse regression model. The reconstructed velocity patterns are very close to the original data. In general, further investigation showed that  parameter identification could be achieved with fewer sensors whereas accurate data reconstruction required more sensors. For example, with even 30 sensors the algorithm was capable of correctly identifying the unknown viscosity (with reduced reconstruction accuracy). Another important observation is the sensitivity of the compressed sensing algorithm to noise. In the presence of noise, the relaxed version of the optimization problem (Eq.~\ref{eqn:mlcs2}) with a relatively high value of $\epsilon$ was required for convergence (herein, $\epsilon$ was set to 25\% of the maximum velocity). In the example considered here, an increase in $\epsilon$ increased reconstruction error as expected but did not affect parameter identification.

\subsection{Opportunities and challenges}
The machine learning ROM framework is capable of merging high-resolution numerical data (with an uncertain parameter) with low-resolution experimental data. A promising application is overcoming limitations in patient-specific CFD and 4D flow MRI characterization of hemodynamics. In CFD modeling, we can generate high-resolution data but often have to face uncertainty in parameters and assumptions. On the other hand, in-vivo 4D flow MRI directly measures blood flow inside the body (without uncertainty in parameters), however, it is subject to imaging artifacts and noise. The machine learning ROM approach could provide a method to overcome these limitations. A similar framework has been used to merge CFD and 4D flow MRI where the authors assumed uncertainty in the CFD inlet flow waveform~\cite{Fathietal18}. The machine learning ROM method could be extended to scenarios where multiple uncertain parameters are present. One challenge in merging CFD and 4D flow MRI data is the discrepancy in geometries during image segmentation and 3D model creation, which could be thought of as another uncertain parameter (the wall location where the no-slip boundary condition is imposed). It might be possible to use the machine learning framework in conjunction with non-uniform rational basis spline (NURBS) surfaces (e.g., isogeometric finite element analysis~\cite{HughesCottrellBazilevs05})  to have control over a set of parameterized geometries. Another challenge is the construction of the training library, which is the computationally expensive part of the model. Greedy algorithms that can appropriately identify the set of parameters in the library construction process could mitigate this difficulty~\cite{Hesthavenetal16}.

\section{Low-rank data recovery from random spatiotemporal measurements  }
\subsection{Motivation} In the previous sections, to reconstruct high-resolution data, we needed prior knowledge of a basis where the data was sparse. Such knowledge might not always be available. However, if the data matrix (Eq.~\ref{eqn:x}) is low-rank, then the full data could be recovered with partial data sampling using low-rank matrix recovery~\cite{majumdar2018compressed}. A low-rank matrix implies that most of the columns of the matrix (herein, temporal snapshots) are not linearly independent. In other words, the total degree of freedom of the matrix is much smaller than the total number of entries, therefore, we may hope to be able to reconstruct the full data with partial observation. In complex cardiovascular flows, given a large number of spatial data points (number of rows) and the complexity in the flow, perfectly correlated columns might not be observed. For instance, noise in data either due to measurement or numerical errors will cause the data matrix to be full rank.  Nevertheless, a high correlation in temporal snapshots is often observed in hemodynamics data. As shown below, the low-rank matrix recovery method could also work under these scenarios. Therefore, given low-resolution random spatial and temporal sampling of hemodynamics data, we might be able to reconstruct the full data. 

\subsection{Model theory and background} Consider the data matrix $\mathbf{X}$ in Eq.~\ref{eqn:x}. Assume we only observe a random subset of entries ($\Omega$) in the matrix (random spatial and temporal sampling of data). Our goal is to reconstruct the entire matrix $\mathbf{X}$ with such observation assuming that $\mathbf{X}$  is low rank. Namely,

 \begin{equation}
\min \textup{rank}(\mathbf{X})  \; \;  \textup{s.t.} \; \;  \mathbf{Y} = M_{\Omega}(\mathbf{X}) \;,
\end{equation}
where $M_{\Omega}$ is a masking operator that selects the entries of $\mathbf{X}$ that have been observed as $\mathbf{Y}$. This optimization problem is computationally very expensive to solve (non-deterministic polynomial hard). However, the problem could be relaxed to a convex optimization problem

 \begin{equation}
\min \lVert \mathbf{X} \rVert_*  \; \;  \textup{s.t.} \; \;  \mathbf{Y} = M_{\Omega}(\mathbf{X}) \;,
\end{equation}
where $\lVert . \rVert_* $ is the nuclear norm, in a similar manner to the ``convexification'' employed for robust PCA in section \ref{sec:RPCA}. This could be solved using the singular value shrinkage algorithm~\cite{majumdar2018compressed}.


\subsection{Example: Reconstruct high-resolution blood flow with random spatiotemporal measurements}
\paragraph{\textit{Problem statement:}} Given low-resolution spatiotemporally random sampling of blood flow velocity data in a cerebral aneurysm \edit{and a coronary artery stenosis,} is it possible to recover the high-resolution data?  

\begin{figure}[h!]
\centering
\includegraphics[scale=0.45]{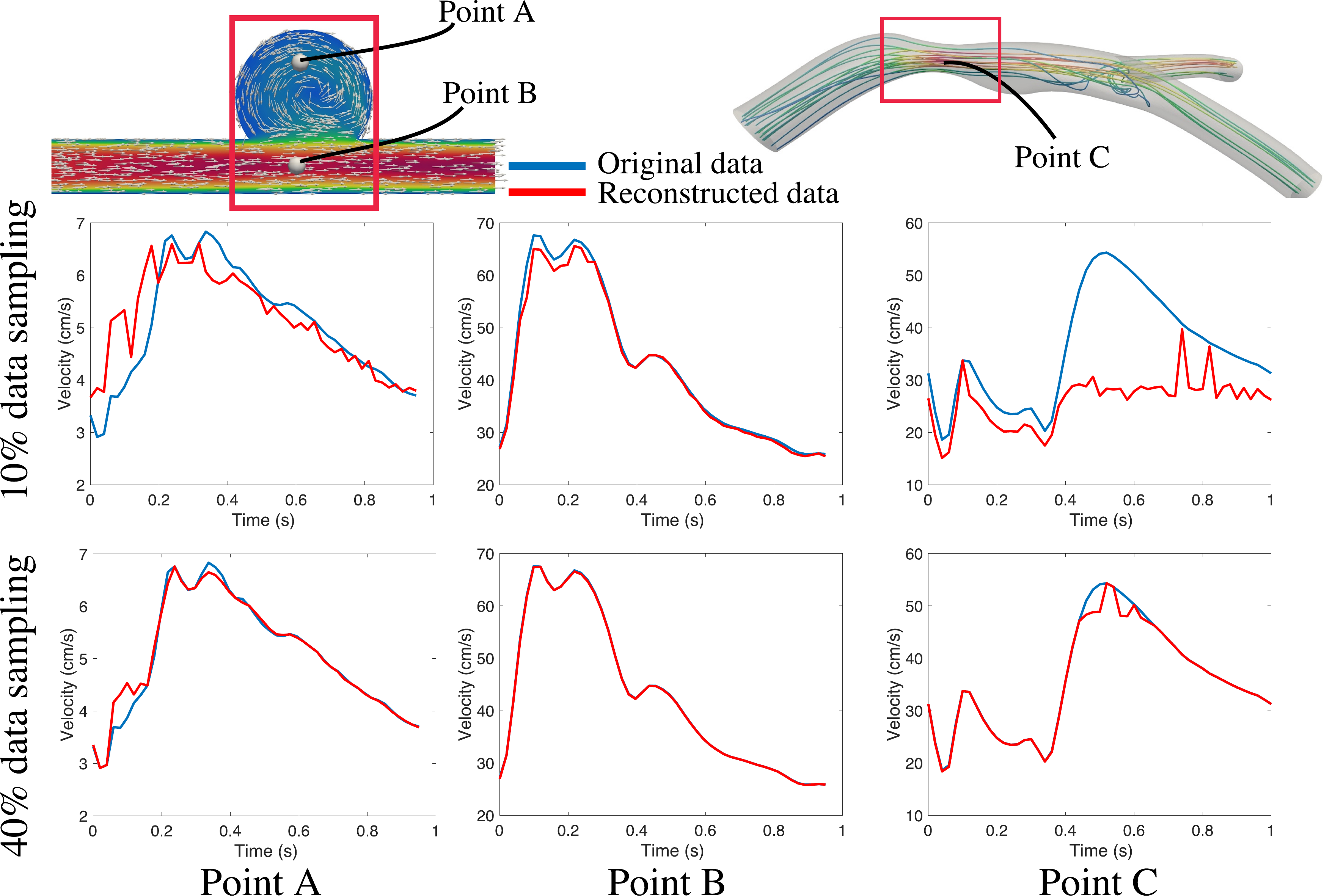}

\caption{\edit{Velocity data is reconstructed in a 2D cerebral aneurysm and a 3D coronary artery stenosis model from random spatiotemporal sampling of the data and using low rank matrix recovery. The boxed region shows the region of interest where the data analysis is performed. The data is reconstructed at point A (inside the aneurysm), point B (inside the parent artery in the aneurysm model), and point C (in the plaque region of the coronary artery model) using 10\% and 40\% random spatiotemporal sampling. The velocity results are plotted versus time where the red line is the reconstructed data and the blue line is the original data.  } }

\label{fig:lowr}
\end{figure}

\paragraph{\textit{Problem solution:}} Consider the pulsatile blood flow problem in the 2D idealized cerebral aneurysm problem in Sec.~\ref{sec:dmde} \edit{and a 3D coronary artery stenosis model from a prior study~\cite{HabibiDawsonArzani20}.} We collect the velocity data in the aneurysmal \edit{and plaque regions} (boxed regions in Fig.~\ref{fig:lowr}) and stack the spatiotemporal velocity data into the data matrix $\mathbf{X}$ similar to Eq.~\ref{eqn:x}. The collected data include vectorial velocity data at 3943 spatial locations and 49 time-steps throughout the cardiac cycle \edit{for the 2D aneurysm model and 73239 spatial locations and 51 intra-cardiac time-steps for the 3D coronary artery model.} Using the low-rank matrix recovery method, we reconstruct the full data given only 10\% and 40\% random spatiotemporal sampling of the data matrix. The results for two points are shown in  Fig.~\ref{fig:lowr}. We observe that 40\% sampling can perfectly capture the velocity outside the aneurysm and with mostly low error for the point inside the aneurysm \edit{and coronary artery stenosis regions.} With 10\% sampling, the algorithm can still perform well for the point outside of the aneurysm with errors near the peak flow rate, whereas the performance for the point inside the aneurysm \edit{and stenosis} where more complex flow is present is not as good.

\subsection{Opportunities and challenges} Matrix completion algorithms provide a promising tool to complete the data when random measurements are missing and a high correlation exists within the data. Similar ideas have been used in reconstructing statistics in turbulent flows~\cite{Zareetal20}. In the context of cardiovascular flows, often the data are correlated, and therefore we may use matrix completion algorithms to increase the spatiotemporal resolution. Interestingly, we observed that the algorithm can successfully complete the data matrix even if the matrix was full-rank, but exhibited low-rank behavior once the threshold used in defining the matrix rank was relaxed. One challenge in such methods is the requirement that the incomplete data sampling should be random (in space and time). However, it might be possible to draw random samples from a uniform sampling to be able to use such methods. Of course, such random samplings need to be carefully devised to avoid accuracy loss.

\section{Sparse identification of nonlinear dynamics (SINDy): Discovering analytical dynamics}
\subsection{Motivation} Extracting analytical governing equations from large datasets has the potential to simplify our physical understanding of dynamical systems and facilitate modeling. Sparse identification of nonlinear dynamics (SINDy) is a recent paradigm that enables the discovery of governing equations from data~\cite{BruntonProctorKutz16}. The central assumption is that the governing equations can be selected sparsely given an appropriately chosen set of candidate functions.  

\subsection{Model theory and background}
Consider a nonlinear dynamical system 
\begin{equation} \label{eqn:dyns}
\dot{\mathbf{x}}(t) = \mathbf{f}(\mathbf{x})  \;,
\end{equation}
where $\mathbf{x}$(t) is the state of the system (e.g., particle position in fluid flow) as a function of time and $\mathbf{f}$ is the dynamics of the system that we would like to approximate analytically and sparsely using a few functions. We consider a library of candidate functions 

\begin{equation}
\Theta (\mathbf{X}) = \begin{bmatrix} 
     \vdots &  \vdots &    \vdots & \vdots &  \vdots &    \vdots &    \vdots   \\
     1  &   	\mathbf{X}	   & 	\mathbf{X}	^{P_2}   & 	\mathbf{X}	^{P_3} & \cdots & \sin(\mathbf{X}) &  \cdots	\\
 \vdots &   \vdots & 	\vdots  & \vdots &   \vdots & 	\vdots &    \vdots 
    \end{bmatrix} \;,
\end{equation}
where $\mathbf{X}^{P_2} $ is the set of quadratic nonlinear functions combining the state vector (e.g., $x_1(t)^2$, $x_1(t)x_2(t)$, ...)  and so forth.  Subsequently, we represent our dynamical system (Eq.~\ref{eqn:dyns}) with these functions

\begin{equation}
\dot{\mathbf{X}}(t)  =  \mathbf{\Theta} (\mathbf{X}) \boldsymbol{\Xi} \;,
\end{equation}
where each column of the matrix $\boldsymbol{\Xi}$, $\boldsymbol{\zeta}_k$ is a sparse vector of coefficients that determines the active functions in $\mathbf{\Theta}$. To find  $\boldsymbol{\zeta}_k$  and therefore the active functions, a convex optimization problem such as the following can be solved

 \begin{equation}
\min_{\boldsymbol{\zeta}_k} \lVert   \dot{\mathbf{x}}(t) -   \mathbf{\Theta} (\mathbf{X}) \boldsymbol{\zeta}_k   \rVert_2  +  \lambda \lVert \boldsymbol{\zeta}_k \rVert_1 \;,
\end{equation}
where the $\lambda$ term promotes sparsity in the selected coefficients. In practice, the identification of the coefficients $\boldsymbol{\zeta}_k$ can instead be solved using a sequential thresholded least-squares algorithm to improve performance under noisy data~\cite{BruntonProctorKutz16}. Finally, the governing equations for the dynamical system could be written for each state $x_k$ as

\begin{equation} 
\dot{x}_k(t) = \mathbf{\Theta} (\mathbf{X})\boldsymbol{\zeta}_k \;.
\end{equation}

\subsection{Example 1: Discover a blood coagulation and thrombosis model  }
\paragraph{\textit{Problem statement:}} Modeling the fluid mechanics of thrombosis (blood clot formation) requires solving large systems of advection-diffusion-reaction equations. It is highly desirable to identify reduced-order thrombosis models from data to simplify thrombosis simulations~\cite{HansenShadden19}. Given the temporal evolution of the prominent biochemicals involved in thrombosis, is it possible to identify the governing equations for the reaction kinetics? 

\begin{figure}[h!]
\centering
\includegraphics[scale=0.34]{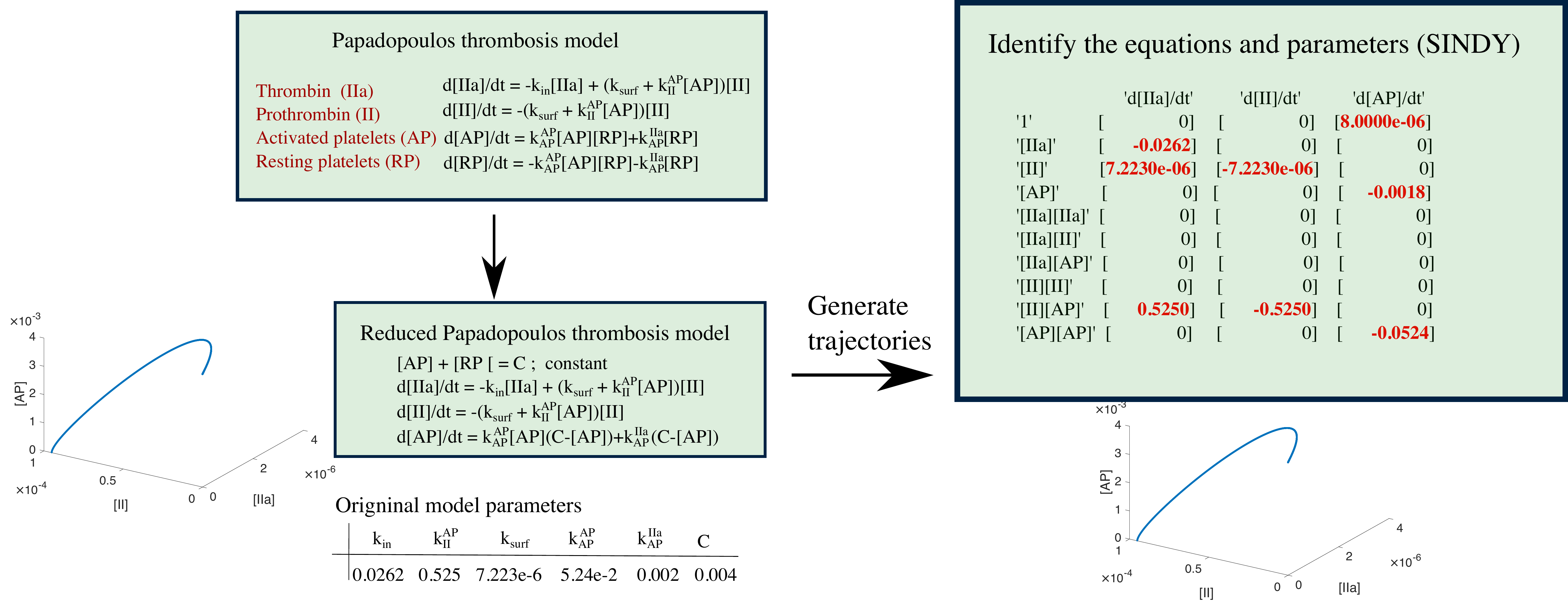}
\caption{ The SINDy algorithm is applied to the Papadopoulos thrombosis model to identify the equations from data. First, a reduced version of the Papadopoulos model is derived by eliminating the activated platelets [AP] using [AP] + [RP] = C where C = 0.004 is determined from the initial conditions ([IIa]$_0$=0, [II]$_0$=9.509e-5,   [AP]$_0$=0,   [RP]$_0$=0.004). The constructed trajectories ([IIa](t), [II](t), and [AP](t)) are used to identify the original governing equations using the sparse regression in SINDy.      }
\label{fig:thro}
\end{figure}

\paragraph{\textit{Problem solution:}} We consider the reduced-order thrombosis model of Papadopoulos~\cite{Papadopoulosetal14,Papadopoulos15} shown in Fig.~\ref{fig:thro}. The system of ordinary differential equations models the biochemical reaction kinetics involved between thrombin (IIa), prothrombin (II), activated platelets (AP), and resting platelets (RP). First, we eliminate the resting platelets from the system by realizing that the [AP] + [RP] is constant. This is an essential step since the SINDy algorithm performs poorly when some variables are highly correlated, due to $\mathbf{\Theta}$ becoming ill-conditioned. The reduced Papadopoulos model is solved to generate biochemical trajectories. In real practice, these trajectories come from experimental data or high-dimensional thrombosis models and the goal is to identify the governing equations for the reaction kinetics based on these trajectories. In constructing the library $\mathbf{\Theta}$, the set of all linear and quadratic functions (based on state variables) are considered. Given the current model, considering a broader set of linearly independent functions did not affect the results. The original model, the generated trajectory, and the identified equations and trajectory from SINDy are shown in Fig.~\ref{fig:thro}. Using the solution trajectories, SINDy exactly captures the active terms and coefficients in the governing equations. 

\subsection{Example 2: Discover analytical velocity in transient vortical flows }
\paragraph{\textit{Problem statement:}} Given particle trajectories in an unsteady 3D vortical flow, is it possible to estimate the time-dependent velocity vector field using SINDY? 

\begin{figure}[h!]
\centering
\includegraphics[scale=0.5]{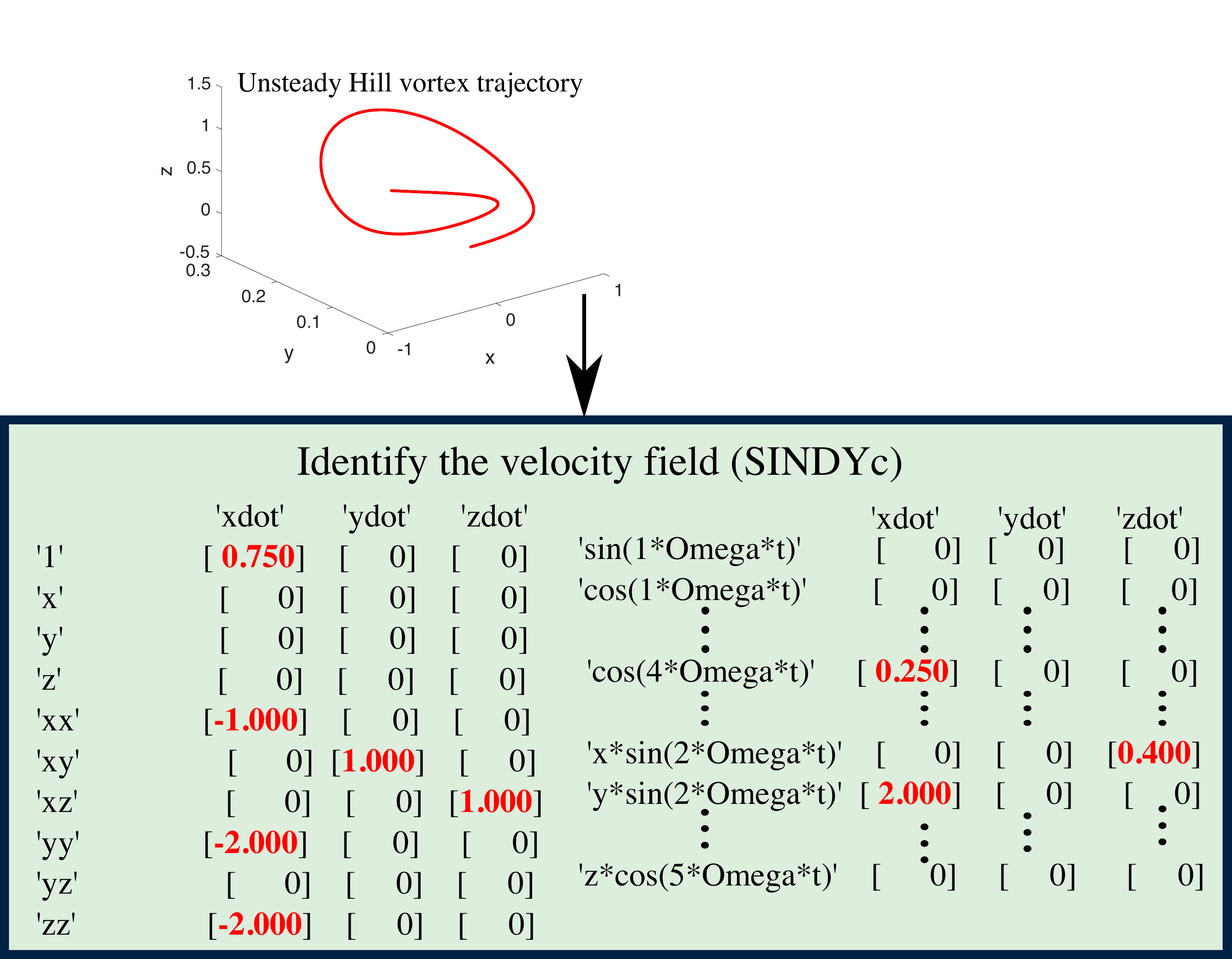}
\caption{ The SINDy with control (SINDYc) algorithm is applied to a particle trajectory generated from an unsteady version of Hill's spherical vortex. The particle trajectory is plotted and the terms identified for the velocity vector field from SINDYc are listed. SINDYc exactly captures the correct terms.  }
\label{fig:sindyc}
\end{figure}

\paragraph{\textit{Problem solution:}} We consider a modified version of the unsteady Hill's spherical vortex in Sec.~\ref{sec:hill} where the 3D velocity field is given by

\begin{equation} \label{eqn:hill2}
\begin{cases}
u(x,y,z,t) =    x^2 + 1 - 2r^2 =  0.75 +  0.25\cos (4\omega t) - x^2 - 2y^2 - 2z^2 + 2y\sin (2\omega t) \\
v(x,y,z,t) =    xy \\
w(x,y,z,t) =    x\big(z + 0.4\sin( 2\omega t) \big) \;,
\end{cases}
\end{equation}
where $\omega = \frac{\pi}{100}$ and  $r=\sqrt{ x^2 + (y-  0.5\sin (2\omega t) )^2 + z^2}$. We simulate particle transport for 8 seconds under this flow for a particle with the initial position of (x$_0$, y$_0$, z$_0$) = (0.02, 0.05, 0.01). Given the resulting trajectory \big(x(t),y(t),z(t)\big), we would like to find an analytical representation for the velocity field. Here, we use SINDy with control (SINDYc)~\cite{BruntonProctorKutz16b} as a simple extension of the SINDy algorithm to be able to estimate time-dependent dynamical systems. For the library $\Theta$, the set of all linear and quadratic functions in x,y,z as well as $\sin( j\omega t)$, $\cos( j\omega t)$, $x_i\sin( j\omega t)$, $x_i\cos( j\omega t)$ are considered where j=1,2,...,5, and $x_i$=x,y,z. For simplicity, we have assumed prior knowledge about $\omega$, however, it is possible to consider a wide range of frequencies by expanding the library. The results are shown in Fig.~\ref{fig:sindyc}. Comparing the active velocity terms and coefficients with Eq.~\ref{eqn:hill2}, we see that the correct terms are accurately predicted.

\subsection{Opportunities and challenges} We have demonstrated two examples of SINDy that could be used in cardiovascular flow modeling. Systems biology models (system of ordinary differential equations governing biological reaction kinetics) are an essential part of multiscale mechanobiology models of disease growth~\cite{ArzaniMastersMofrad17,SreeTepole20}. SINDy provides a framework to derive such models from experimental data or identify reduced-order systems biology models from higher-order models. Identifying analytical velocity fields from particle trajectories is another example. Of course, this is expected to be a challenging task in complex cardiovascular flows. We have shown an example based on an idealized vortex flow. It remains to be investigated if similar models could be derived from other vortex-dominated flows such as blood flow in the left ventricle. One challenge in the application of SINDY is that the matrix $\Theta$ becomes ill-conditioned once the trajectories are highly correlated, which will compromise the algorithm results. A reweighted $l_{1}$ regularized least squares technique has been proposed to overcome this issue~\cite{Cortiellaetal20}. Another challenge is in building the library of candidate functions. In the examples considered, not considering an important active term in the library, resulted in a dense solution where most terms became active. It should be noted that depending on the complexity of the system studied, it may not always be possible to represent a dynamical system sparsely. Finally, SINDY has been extended to partial differential equations (PDE) to identify active terms and coefficients in spatiotemporal PDE models~\cite{Rudyetal17}. 

%
%
%
%

\section{Other techniques and applications}
In this section, we briefly discuss some of the recent trends in data-driven modeling and data science and their potential in cardiovascular flow modeling. 

\paragraph{Machine learning and neural networks} Neural networks are a powerful class of machine learning techniques and may be thought of as an approximator to complex nonlinear functions. A popular application in cardiovascular research is learning clinical outcomes based on different input parameters derived from patient-specific simulations~\cite{Detmeretal20,Jiangetal20}. We may think of this as an advanced multi-parameter regression framework where we have several input parameters, different clinical outcomes, and large databases where we are interested in learning/fitting a statistical relationship between the data, which is achieved through the learning process. Another application area is learning models from complex large datasets. An active area of research related to this is learning turbulence models based on direct numerical simulation (DNS) databases~\cite{Duraisamyetal19}. Given the unique physics of turbulence in cardiovascular flows~\cite{AntigaSteinma09,Xuetal20}, learning customized Reynolds averaged Navier-Stokes (RANS) or large-eddy simulation (LES) models based on high-fidelity DNS data of turbulent pulsatile blood flow seems to be a promising area of future research. A common question for machine learning is if it can replace physics-based cardiovascular flow simulations, and therefore provide a very fast patient-specific estimation of hemodynamics once the one-time expensive training process is completed. Currently, this paradigm has been successfully used to estimate CFD-based fractional flow reserve (FFR) in coronary artery stenosis~\cite{Coenenetal18}. However, FFR is a single variable that measures the reduction in flow rate (pressure drop) due to a stenosis. Using a similar paradigm for estimating variables that exhibit spatial and temporal variation (e.g., velocity or wall shear stress)~\cite{Sarramietal16} is a much more involved process, in which extremely large datasets are required, and it is not clear if it could be achieved in the near future. In fact, even large datasets are likely not ``large'' enough when it comes to systems that are highly sensitive to variations in input data~\cite{SucciCoveney19}. The recent physics-informed neural networks (PINN) paradigm~\cite{Raissietal19} considers the governing equations in the learning process and could ultimately reduce some of the costs associated with traditional CFD modeling such as mesh creation. PINN has been used for an example inverse modeling problem in cardiovascular flows where one is interested in estimating the velocity field based on concentration data~\cite{Raissietal20}. This approach seems to pave the way to a new in-vivo framework for measuring velocity based on time-resolved concentration measured using medical imaging (e.g., dynamic contrast-enhanced imaging). \edit{Finally, PINNs have been recently used for improving 4D flow MRI data fidelity (super-resolution and denoising)~\cite{Fathietal20}.}

\paragraph{Parametric reduced-order models} The reduced-order models that we discussed in Sec.~\ref{sec:dmd} did not consider parameter variation. Namely, we approximated them given spatiotemporal velocity data for a fixed parameter. Sometimes in computational modeling, we are interested in a range of parameters. Reduced-order modeling tools such as the Galerkin-POD learn a projected space and basis, which they  use to evolve the solution and therefore provide a computationally efficient prediction of the solution. These methods are trained based on fixed parameters and therefore cannot be used when parameters in the model change. Parametric reduced-order models have been developed to address this challenge~\cite{BennerGugercinWilcox15,Hesthavenetal16}. These reduced-order modeling techniques could be valuable in multiscale modeling of cardiovascular disease growth where parameters in the blood flow or structural mechanics models can change during disease growth simulation.

\paragraph{Data-driven multiscale modeling} Multiscale modeling of cardiovascular disease is challenging as such models need to connect different models across different spatial and temporal scales. Data-driven modeling and machine learning have the potential to transform multiscale modeling of cardiovascular disease~\cite{Pengetal20}. Reduced-order models can be used to accelerate computationally expensive multiscale models. We may use the data generated from expensive multiscale simulations or experimental observations to derive a simplified approximation of the underlying physics (e.g., using SINDy). Complete collection of biological data across multiple scales represents a major challenge in calibrating parameters in multiscale models. Developing data-driven multiscale models that can leverage sparsity or are robust to corrupt and incomplete data is a promising topic of future research. In prior sections, we discussed data-driven modeling with multi-modality data. Developing multiscale models that can appropriately leverage multi-modality data is another interesting topic~\cite{Perdikarisetal16}. Deep neural networks provide a promising framework for multiphysics and multiscale modeling~\cite{Caietal20}. Combining data-driven modeling and machine learning with physics-based simulations provides a hybrid modeling strategy~\cite{Vonetal20} with high potential in multiscale modeling of biological systems~\cite{Pengetal20}. Advancements in each of these fields and their coupling will produce predictive digital twins~\cite{KapteynWillcox20} that could facilitate treatment planning, patient management, and ultimately transform personalized cardiovascular medicine~\cite{Hoseetal19}.


\section{Conclusion}
Recent advances in computational power, data science, and abundance of data are starting to revolutionize different fields of science and engineering. Data-driven cardiovascular flow modeling is currently at its infancy, yet there are various opportunities to develop customized data-driven models that can transform cardiovascular biomechanics research. Multi-modality and multi-fidelity data assimilation, overcoming poor data quality, parameter identification, super-resolution flow reconstruction, and reduced-order modeling are some examples. We hope that this manuscript inspires researchers to develop data-driven models that can address outstanding challenges in the field of cardiovascular biomechanics.

\section*{Conflict of interest}
The authors declare no conflict of interest.

\section*{Acknowledgement}
A. A. acknowledges funding from NSF OAC grant No.~1947559. The authors would like to thank Milad Habibi for assistance in generating the POD and DMD data in Sec~\ref{sec:dmd} and Dr. Roshan M. D'Souza for fruitful discussions related to the topic of the manuscript. We are also thankful to Dr. Steven Brunton and Dr. Nathan Kutz for permission to use their codes in several examples in this manuscript. 

\section*{Data Availability}
\edit{The codes and data used to generate the results in the manuscript are available on Github \url{https://github.com/amir-cardiolab/cardio-data-driven}.  }


\bibliographystyle{unsrt}

\end{document}